\documentclass[journal=jpccck,manuscript=article]{achemso}

\usepackage[T1]{fontenc}
\usepackage{amsthm,amsmath,amssymb}
\usepackage{graphicx}
\graphicspath{{figures/}}
\usepackage[unicode=true,breaklinks=true,colorlinks=true,linkcolor=blue,citecolor=blue,urlcolor=blue]{hyperref}

\author{Jiayue Han}
\affiliation{Department of Chemistry, School of Science and Research Center for Industries of the Future, Westlake University, Hangzhou, Zhejiang 310030, China}

\author{Yu Wang}
\email{yu_wang@hebut.edu.cn}
\affiliation{School of Chemical Engineering, Hebei University of Technology, Tianjin 300401, China}

\author{Vahid Mosallanejad}
\affiliation{Department of Chemistry, School of Science and Research Center for Industries of the Future, Westlake University, Hangzhou, Zhejiang 310030, China}
\alsoaffiliation{Institute of Natural Sciences, Westlake Institute for Advanced Study, Hangzhou, Zhejiang 310024, China}

\author{Wei Liu}
\affiliation{Department of Chemistry, School of Science and Research Center for Industries of the Future, Westlake University, Hangzhou, Zhejiang 310030, China}

\author{Wenjie Dou}
\email{douwenjie@westlake.edu.cn}
\affiliation{Department of Chemistry, School of Science and Research Center for Industries of the Future, Westlake University, Hangzhou, Zhejiang 310030, China}
\alsoaffiliation{Institute of Natural Sciences, Westlake Institute for Advanced Study, Hangzhou, Zhejiang 310024, China}
\alsoaffiliation{Key Laboratory for Quantum Materials of Zhejiang Province, Department of Physics, School of Science and Research Center for Industries of the Future, Westlake University, Hangzhou, Zhejiang 310030, China}

\title{Floquet Nonadiabatic Dynamics for Light--Matter Interactions: Recent Advances and Emerging Opportunities}

\begin{document}

\begin{abstract}
Light--matter interactions provide versatile routes for probing and controlling chemical reactivity, charge transport, and material properties. Time-periodic external fields can reshape electronic states and open new dynamical pathways beyond the field-free Born--Oppenheimer (BO) picture. Floquet nonadiabatic dynamics has consequently emerged as an important framework for describing coupled electron--nuclear dynamics under periodic driving. In this Perspective, we first discuss recent developments in Floquet nonadiabatic dynamics methods for closed and open quantum systems. We then highlight how this framework provides mechanistic insights into electron transfer at molecule--metal interfaces, quantum transport in molecular junctions, carrier dynamics in crystalline solids, and multicolor Floquet engineering. Finally, we outline key conceptual and computational challenges that must be addressed to transform Floquet nonadiabatic dynamics from model-based demonstrations into predictive, first-principles simulations of realistic light-driven processes.
\end{abstract}

\begin{figure}[hb]
\centering
\includegraphics[width=0.60\linewidth]{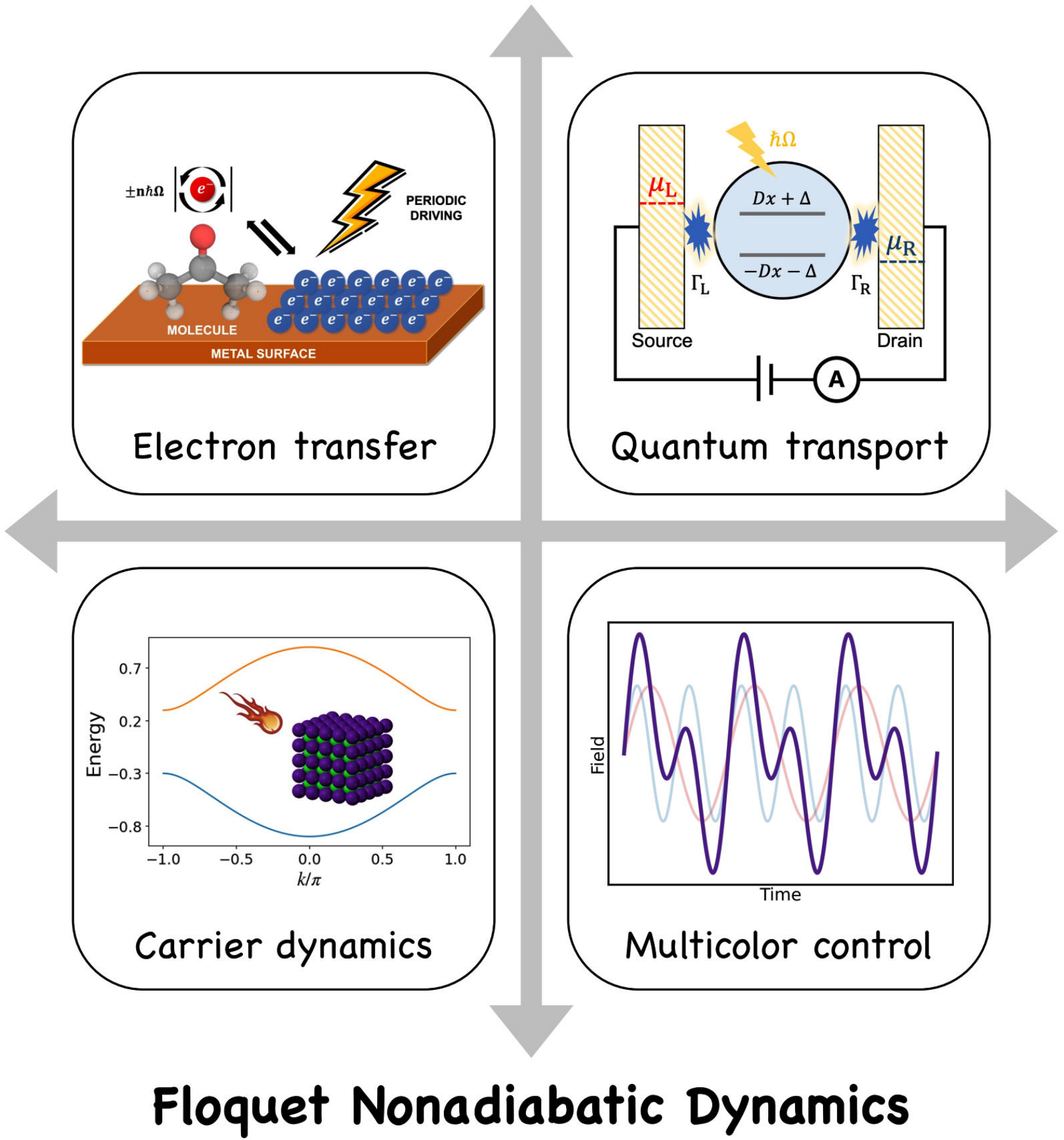}
\end{figure}

\newpage

\section{Introduction}

Light--matter interactions underpin a wide range of phenomena in molecular, interfacial, and condensed-phase systems.\cite{Hertzog2019strong,Weight2023theory,Wu2025light,Nitzan2024chemical} This versatility arises because external electromagnetic fields provide tunable control parameters, such as frequency, amplitude, phase, and polarization, that can modulate molecular and material properties without direct chemical modification.\cite{Garcia_Vidal2021manipulating,Svendsen2024ab} In theoretical modeling, external electromagnetic fields are typically described in two complementary ways.\cite{Ahmed2024a} First, the field can be treated as a classical electromagnetic wave that drives matter out of equilibrium, which is widely used in areas such as ultrafast spectroscopy,\cite{Thoss2007correlated,Ibrahim2018h2,Maiuri2019ultrafast} photocatalysis,\cite{Zhang2017heterostructures,Zhao2020two} strong-field coherent control,\cite{Bandrauk1981photodissociation,Corrales2014control,Corrales2017strong} plasmon-driven chemistry,\cite{Zhang2017surface,Zhang2019plasmon,Sarkar2025plasmon} and light-induced energy and charge transport.\cite{Tikhonov2002calculating,KOHLER2005driven,Zhang2024enhanced} Second, in contexts such as cavity quantum electrodynamics (cavity QED)\cite{Walther2006cavity,Taylor2025light} and polaritonic chemistry,\cite{Feist2017polaritonic,Ribeiro2018polariton,Xiang2020intermolecular} selected electromagnetic modes are commonly treated quantum mechanically. These quantized modes can hybridize with matter excitations to form polaritons,\cite{Rivera2020light,Xie2025exploiting} thereby reshaping potential energy landscapes\cite{Thomas2019tilting} and modifying reaction pathways.\cite{Coles2011vibrationally,Coles2013imaging}

Motivated by these advances, considerable attention has been devoted to the theoretical description of molecular dynamics driven by time-periodic classical fields.\cite{Zhang2019non} Conventional molecular dynamics simulations are largely based on the Born--Oppenheimer (BO) approximation, which assumes that nuclei evolve on a single potential energy surface (PES) and the electronic degrees of freedom (DOFs) adjust instantaneously to nuclear motion.\cite{Ess_n1977the,Wudka1990remarks} Although the BO approximation has proven successful for many isolated or weakly perturbed molecular systems, it can break down when electronic and nuclear motions become strongly coupled.\cite{Butler1998chemical,Worth2004beyond} This naturally brings the problem into the realm of nonadiabatic dynamics.\cite{Nitzan2024chemical,Subotnik2016understanding} For open quantum systems, the theory is further complicated by system--environment coupling, electronic relaxation, and dissipation, all of which must be incorporated consistently with the nonadiabatic treatment.\cite{Dou2020nonadiabatic} Periodic driving, which is naturally described using Floquet theory, introduces an additional layer of complexity by opening transition pathways that are absent in field-free dynamics.\cite{Ivanov2021floquet} A central theoretical challenge is therefore to develop a nonperturbative yet tractable framework for incorporating periodic driving into coupled electron--nuclear dynamics in both closed and open quantum systems. These considerations motivate the combination of nonadiabatic dynamics with Floquet theory, namely Floquet nonadiabatic dynamics, which is the central theme of this Perspective.

For systems with a limited number of DOFs, several numerically exact methods are available, including the full electron--nuclear time-dependent Schr\"odinger equation (en-TDSE),\cite{Deumens1994time,Suzuki2014electronic} multi-configuration time-dependent Hartree (MCTDH),\cite{Beck2000the,Wang2003multilayer} hierarchical quantum master equation (HQME),\cite{Schinabeck2016hierarchical,Erpenbeck2019hierarchical} numerical renormalization group (NRG),\cite{Konik2007numerical,Bulla2008numerical} and quantum Monte Carlo (QMC).\cite{M_hlbacher2008real,Weiss2008iterative,Austin2011quantum} These methods can capture nonadiabatic effects with high accuracy, but their computational cost typically grows rapidly with the number of DOFs, limiting their direct application to realistic systems. Therefore, a variety of lower-cost mixed quantum--classical (MQC) methods have been developed, including mean-field (MF) dynamics, surface hopping (SH), and \textit{ab initio} multiple spawning (AIMS).\cite{Tully1998mixed,Crespo_Otero2018recent} By treating electronic DOFs quantum mechanically while propagating nuclear motion classically, MQC methods provide a practical compromise between accuracy and computational efficiency. Recent extensions of MQC dynamics have provided essential foundations for the Floquet nonadiabatic dynamics methods discussed in this Perspective. For example, Dou and co-workers have developed nonadiabatic dynamics methods for molecules at metal surfaces across different molecule--metal coupling regimes.\cite{Dou2016a,Dou2017a,Dou2016electronic,Miao2017vibrational} Krotz and co-workers formulated reciprocal-space MQC dynamics, enabling efficient simulations of carrier dynamics in periodic crystalline materials.\cite{Krotz2021a,Krotz2022a}

Floquet theory is a natural choice for treating external periodic driving in nonadiabatic dynamics.\cite{Floquet1883sur,Shirley1965solution,Ivanov2021floquet} Analogous to Bloch's theorem for spatially periodic crystals, Floquet theory enables a time-periodic Hamiltonian to be recast into a time-independent form within an extended Hilbert space.\cite{Traversa2013generalized,Mosallanejad2023floquet,Mosallanejad2025floquet} Combining Floquet theory with MQC dynamics yields trajectory-based methods that can describe photon-dressed electronic states, account for photon-assisted nonadiabatic transitions, and incorporate light-induced modifications of forces and couplings.\cite{Zhou2020nonadiabatic,Zhou2023nonadiabatic} When system--bath coupling is further incorporated, these methods can also describe driven relaxation and nonequilibrium steady states in dissipative environments.\cite{Wang2023nonadiabatic_FSH,Wang2023nonadiabatic_FEF,Wang2024nonadiabatic} Most Floquet nonadiabatic dynamics studies to date have focused on one-frequency driving. Although two-mode Floquet theory has been developed over several decades,\cite{Ho1983semiclassical1,Ho1984semiclassical2,Ho1985semiclassical3,Ho1985semiclassical4,Son2008many,Poertner2020validity,Gustin2021high,Barriga2024floquet,Duha2024two} its integration with nonadiabatic dynamics remains challenging. Recent progress in two-mode Floquet nonadiabatic dynamics will be discussed in later sections.

The present Perspective is not intended to provide an exhaustive review of Floquet nonadiabatic dynamics. Rather, we aim to present our perspective on this emerging field and summarize lessons learned from our investigations over the past few years. Section~\nameref{sec:theory} introduces a minimal theoretical framework for Floquet nonadiabatic dynamics. We then discuss four interconnected application areas: light-driven electron transfer at molecule--metal interfaces, light-modulated charge and spin transport in molecular junctions, Floquet carrier dynamics in crystalline solids, and two-mode Floquet engineering with bichromatic fields. Section~\nameref{sec:challenges_and_opportunities} outlines key challenges and opportunities, and Section~\nameref{sec:summary} concludes with a brief summary.

\begin{figure}[hb]
\centering
\includegraphics[width=\linewidth]{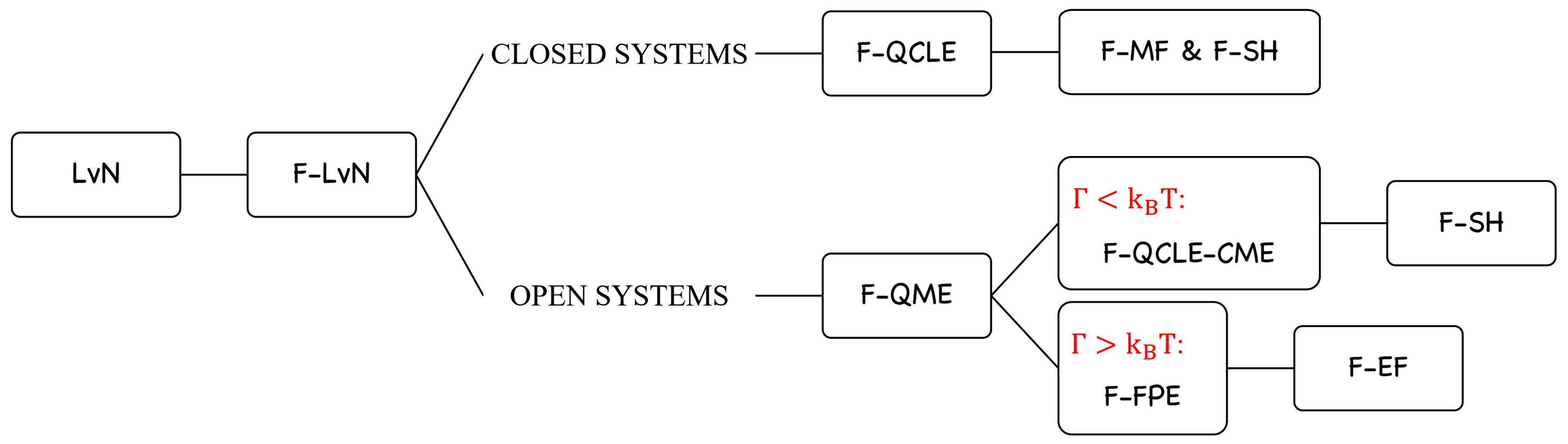}
\caption{Schematic overview of the minimal theoretical framework for Floquet nonadiabatic dynamics discussed in this Perspective. Starting from the Liouville--von Neumann (LvN) equation, one obtains the Floquet Liouville--von Neumann (F-LvN) equation. For closed systems, a partial Wigner transform leads to the Floquet quantum--classical Liouville equation (F-QCLE), which can be solved by Floquet mean-field (F-MF) dynamics and Floquet surface hopping (F-SH). For open systems, the Floquet quantum master equation (F-QME) leads to the Floquet quantum--classical Liouville equation--classical master equation (F-QCLE-CME) and F-SH in the weak-coupling regime ($\Gamma < k_{\mathrm{B}}T$), or to the Floquet Fokker--Planck equation (F-FPE) and Floquet electronic friction (F-EF) in the strong-coupling regime ($\Gamma > k_{\mathrm{B}}T$).}
\label{fig:theory_flowchart}
\end{figure}

\section{Minimal Theoretical Framework for Floquet Nonadiabatic Dynamics}
\label{sec:theory}

We begin with the Liouville--von Neumann (LvN) equation:
\begin{equation}
\frac{\partial}{\partial t}\hat{\rho}(t) = -\frac{i}{\hbar}\bigl[\hat{H}(t),\hat{\rho}(t)\bigr],
\label{eq:lvn}
\end{equation}
where $\hat{\rho}(t)$ denotes the density operator of the full system and $\hat{H}(t)$ denotes the total Hamiltonian. In the presence of periodic driving, the Hamiltonian satisfies $\hat{H}(t+T)=\hat{H}(t)$, where $T$ is the driving period and $\Omega=2\pi/T$ is the driving frequency. By exploiting Floquet theory, Equation~\eqref{eq:lvn} can be recast into the Floquet Liouville--von Neumann (F-LvN) equation, which serves as a common starting point for the closed- and open-system formulations discussed below.\cite{Mosallanejad2023floquet}

For closed systems, applying a partial Wigner transform to the F-LvN equation with respect to the nuclear DOFs yields the Floquet quantum--classical Liouville equation (F-QCLE), from which trajectory-based MQC methods can be derived.\cite{Horenko2001a,Chen2020on,Mosallanejad2023floquet} Here, we focus on two such methods, Floquet mean-field (F-MF) dynamics and Floquet surface hopping (F-SH), both of which can be formulated in real and reciprocal space.\cite{Chen2024floquet_solid,Wang2024interband}

For open systems, we focus primarily on driven molecular subsystems coupled to metallic electronic continua. In this setting, the reduced dynamics can first be described using a Floquet quantum master equation (F-QME). The F-QME provides a practical and tractable description of dissipative Floquet dynamics, with its range of validity and numerical robustness determined by the Born--Markov approximation.\cite{Wang2023nonadiabatic_FSH} To connect the F-QME to MQC methods, a partial Wigner transform is performed with respect to the nuclear DOFs. Throughout this Perspective, $\Gamma$ denotes the hybridization-induced electronic level broadening, which characterizes the molecule--metal coupling strength, and $k_{\mathrm{B}}T$ denotes the thermal energy of the electronic bath. Depending on whether $\Gamma$ is smaller or larger than $k_{\mathrm{B}}T$, distinct reduced descriptions can be obtained for the weak- and strong-coupling limits. In the weak-coupling regime ($\Gamma < k_{\mathrm{B}}T$), the partial Wigner transform leads to a Floquet quantum--classical Liouville equation--classical master equation (F-QCLE-CME), which can be solved by trajectory-based F-SH.\cite{Wang2024nonadiabatic} In contrast, in the strong-coupling regime ($\Gamma > k_{\mathrm{B}}T$), a velocity expansion of the fast electronic response leads to a Floquet Fokker--Planck equation (F-FPE), which can be solved by trajectory-based Floquet electronic friction (F-EF).\cite{Mosallanejad2023floquet,Wang2023nonadiabatic_FEF} Figure~\ref{fig:theory_flowchart} summarizes the connections among the Floquet nonadiabatic dynamics methods discussed in this Perspective.

\subsection{Floquet Liouville--von Neumann Equation}

As introduced above, the equation of motion for the total density operator in the Schr\"odinger picture is given by Equation~\eqref{eq:lvn}. To derive the Floquet representation of the LvN equation, we first expand both the Hamiltonian and the density operator in a discrete Fourier series
\begin{equation}
\hat{H}(t) = \sum_{n\in\mathbb{Z}}\hat{H}^{n} e^{in\Omega t}, 
\qquad
\hat{\rho}(t) = \sum_{n\in\mathbb{Z}}\hat{\rho}^{n}(t) e^{in\Omega t},
\label{eq:fourier-expansion}
\end{equation}
where $\hat{H}^{n}$ denotes the time-independent Fourier component of the periodic Hamiltonian, and $\hat{\rho}^{n}(t)$ denotes the corresponding time-dependent coefficient of the density operator.

We next embed the original time-periodic problem into an extended Floquet space,
\begin{equation}
\hat{H}^F =
\sum_{n\in\mathbb{Z}}\hat{L}_n \otimes \hat{H}^{n}
+
\hat{N} \otimes \hat{I}\,\hbar\Omega,
\qquad
\hat{\rho}^F(t) =
\sum_{n\in\mathbb{Z}}\hat{L}_n \otimes \hat{\rho}^{n}(t).
\label{eq:HF_rhoF}
\end{equation}
Here, $\hat{N}$ is the Floquet number operator, $\hat{L}_n$ is the Floquet ladder operator, and $\hat{I}$ is the identity operator in the physical Hilbert space. Substituting Equations~\eqref{eq:HF_rhoF} into Equation~\eqref{eq:lvn} leads to the F-LvN equation:
\begin{equation}
\frac{\partial}{\partial t}\hat{\rho}^F(t) = -\frac{i}{\hbar}\bigl[\hat{H}^F,\hat{\rho}^F(t)\bigr].
\label{eq:flvn}
\end{equation}
Equation~\eqref{eq:flvn} has the same formal structure as the original LvN equation, but it now involves the time-independent Floquet Hamiltonian $\hat{H}^F$ in the extended Floquet space. As discussed in Section~\nameref{sec:two-frequency}, the two-mode Floquet formalism leads to an F-LvN equation with the same formal structure as its single-mode counterpart.

\subsection{Closed-System Floquet Nonadiabatic Dynamics}

For a closed system, the Floquet Hamiltonian $\hat{H}^F$ describes the periodically driven target system itself, with no explicit coupling to an external environment. For an electron--nuclear system, one commonly partitions the Hamiltonian as
\begin{equation}
\hat{H}^{F}
=
\hat{T}_{\mathrm{nuc}}
+
\hat{H}^{F}_{\mathrm{el}}(\hat{\mathbf{R}}).
\label{eq:HF_closed_decomp}
\end{equation}
Here, $\hat{T}_{\mathrm{nuc}}$ is the nuclear kinetic-energy operator, and $\hat{H}^{F}_{\mathrm{el}}(\hat{\mathbf{R}})$ is the Floquet electronic Hamiltonian, including the nuclear-coordinate-dependent electronic and interaction terms.

\subsubsection{Floquet Quantum--Classical Liouville Equation}

Applying a partial Wigner transform to the F-LvN equation in Equation~\eqref{eq:flvn} with respect to the nuclear DOFs gives a phase-space representation of the coupled electron--nuclear dynamics. Keeping the leading-order terms in the Moyal expansion yields the F-QCLE:
\begin{equation}
\begin{aligned}
\frac{\partial}{\partial t}\hat{\rho}_{W}^F(\mathbf{R},\mathbf{P},t)
={}&
\frac{1}{2}\bigl\{\hat{H}_{W}^F(\mathbf{R},\mathbf{P}),\hat{\rho}_{W}^F(\mathbf{R},\mathbf{P},t)\bigr\}
- \frac{1}{2}\bigl\{\hat{\rho}_{W}^F(\mathbf{R},\mathbf{P},t),\hat{H}_{W}^F(\mathbf{R},\mathbf{P})\bigr\}
\\
&{}- \frac{i}{\hbar}\bigl[\hat{H}_{W}^F(\mathbf{R},\mathbf{P}),\hat{\rho}_{W}^F(\mathbf{R},\mathbf{P},t)\bigr].
\end{aligned}
\label{eq:F-QCLE}
\end{equation}
Here, the subscript $W$ denotes a partially Wigner-transformed quantity, and the curly brackets denote Poisson brackets with respect to the classical variables $(\mathbf{R},\mathbf{P})$, defined as
\begin{equation}
\{A,B\}
=
\sum_{\alpha}
\left(
\frac{\partial A}{\partial R_{\alpha}}
\frac{\partial B}{\partial P_{\alpha}}
-
\frac{\partial A}{\partial P_{\alpha}}
\frac{\partial B}{\partial R_{\alpha}}
\right).
\label{eq:poisson_bracket}
\end{equation}
The F-QCLE motivates trajectory-based approximations such as F-MF dynamics and F-SH. As discussed in Section~\nameref{sec:k-space}, both methods can also be formulated in reciprocal space.

\subsubsection{Floquet Mean-Field Dynamics}

In F-MF dynamics, classical trajectories evolve under an averaged force generated by the Floquet electronic states.\cite{Ehrenfest1927bemerkung} For classical nuclear coordinates $\mathbf{R}$ and conjugate momenta $\mathbf{P}$, the equations of motion can be written as
\begin{equation}
\begin{aligned}
\dot{\mathbf{R}}
={}&
\frac{\partial T_{\mathrm{nuc}}(\mathbf{P})}
{\partial \mathbf{P}},
\\[0.5em]
\dot{\mathbf{P}}
={}&
-
\mathrm{Tr}
\left[
\frac{\partial \hat{H}^{F}_{\mathrm{el}}(\mathbf{R})}
{\partial \mathbf{R}}
\hat{\rho}^{F}
\right].
\end{aligned}
\label{eq:fmf_real}
\end{equation}
Here, $T_{\mathrm{nuc}}(\mathbf{P})$ is the classical nuclear kinetic energy, and $\hat{H}^{F}_{\mathrm{el}}(\mathbf{R})$ is the Floquet electronic Hamiltonian.

\subsubsection{Floquet Surface Hopping}

In F-SH, each trajectory evolves on one active Floquet quasi-energy surface, with stochastic hops between surfaces determined by Floquet nonadiabatic couplings.\cite{Tully1990molecular} The hopping rate from surface $N$ to $M$ can be written as
\begin{equation}
k_{N\rightarrow M}
=
\max
\left\{
0,
-2\,\mathrm{Re}
\left[
\dot{\mathbf{R}}\cdot
\mathbf{d}_{MN}^{F}(\mathbf{R})
\frac{\sigma_{NM}^{F(\mathrm{ad})}}{\sigma_{NN}^{F(\mathrm{ad})}}
\right]
\right\}.
\label{eq:fsh_real}
\end{equation}
Here, $\sigma^{F(\mathrm{ad})}$ is the Floquet electronic density matrix in the adiabatic representation, and $\mathbf{d}_{MN}^{F}(\mathbf{R})$ is the Floquet derivative coupling between states $M$ and $N$. After a hop, the classical momentum is rescaled along the Floquet derivative-coupling direction to conserve the total energy.

\subsection{Open-System Floquet Nonadiabatic Dynamics}

Having recast the full dynamics into the F-LvN form in Equation~\eqref{eq:flvn}, we now consider an open system in which a driven molecular subsystem is coupled to a metallic electronic bath. As in standard open-system theory, the total Floquet Hamiltonian can be partitioned into system, bath, and system--bath coupling terms:\cite{Mosallanejad2025floquet}
\begin{equation}
\hat{H}^F = \hat{H}_s^F + \hat{H}_b^F + \hat{H}_{sb}^F,
\label{eq:HF-partition}
\end{equation}
where $\hat{H}_s^F$ acts on the molecular subsystem, $\hat{H}_b^F$ describes the metallic bath, and $\hat{H}_{sb}^F$ encodes the system--bath coupling. Consistent with the closed-system discussion, $\hat{H}_s^F$ contains the driven nuclear and electronic DOFs of the molecular subsystem, including their mutual coupling.

\subsubsection{Floquet Quantum Master Equation}

We focus on the reduced Floquet density operator of the molecular subsystem, $\hat{\rho}_s^F(t)=\mathrm{Tr}_b[\hat{\rho}^F(t)]$, obtained by tracing the total Floquet density operator over the bath DOFs. To derive a closed equation of motion for the reduced Floquet density operator, we employ the Born--Markov approximation: system--bath correlations are assumed to remain weak, and the metallic bath is treated as a stationary grand-canonical reservoir characterized by its chemical potential and temperature. The resulting reduced dynamics can be written as an F-QME:
\begin{equation}
\frac{\partial}{\partial t}\hat{\rho}_s^F(t)
=
-\frac{i}{\hbar}\bigl[\hat{H}_s^F,\hat{\rho}_s^F(t)\bigr]
- \hat{\hat{\mathcal{L}}}_{sb}^F \hat{\rho}_s^F(t).
\label{eq:F-QME}
\end{equation}
Here and below, double-hatted calligraphic symbols such as $\hat{\hat{\mathcal{L}}}_{sb}^F$ denote superoperators that act linearly on system operators. The Redfield-type dissipator in Equation~\eqref{eq:F-QME} takes the form
\begin{equation}
\begin{aligned}
\hat{\hat{\mathcal{L}}}_{sb}^F \hat{\rho}_s^F(t)
={}&
\frac{1}{\hbar^2}
\int_{0}^{\infty}\! \mathrm{d}\tau\;
e^{-\tfrac{i}{\hbar}\hat{H}_s^F t}
\mathrm{Tr}_b
\Bigl[
    \hat{H}_{sb,I}^F(t),
    \Bigl[
        \hat{H}_{sb,I}^F(t-\tau),
        \\
        &e^{\tfrac{i}{\hbar}\hat{H}_s^F t}
        \hat{\rho}_s^F(t)
        e^{-\tfrac{i}{\hbar}\hat{H}_s^F t}
        \otimes
        \hat{\rho}_b^{\mathrm{eq}}
    \Bigr]
\Bigr]
e^{\tfrac{i}{\hbar}\hat{H}_s^F t},
\end{aligned}
\label{eq:LsbF}
\end{equation}
where $\hat{H}_{sb,I}^F(t)$ denotes the system--bath coupling in the interaction picture, and $\hat{\rho}_b^{\mathrm{eq}}$ denotes the equilibrium density operator of the bath.

This formulation treats the molecular subsystem quantum mechanically in the Floquet representation while incorporating dissipation induced by the metallic bath. For long-time dynamics with explicit nuclear motion, however, an MQC treatment is often more practical. We therefore next introduce MQC reduced descriptions obtained through a partial Wigner transform in the weak- and strong-coupling regimes.

\subsubsection{Floquet Mixed Quantum--Classical Dynamics in the Weak-Coupling Regime}

In the weak-coupling regime ($\Gamma < k_{\mathrm{B}}T$), molecule--metal hybridization is sufficiently small that discrete molecular states remain only weakly broadened by the metallic electronic continuum. Nonadiabatic electronic transitions between molecular electronic states and metallic electronic states can then be treated perturbatively, and the Born--Markov approximation underlying the F-QME is well justified. In this limit, we arrive at the F-QCLE-CME and its efficient trajectory-based realization, F-SH.

\paragraph{Floquet Quantum--Classical Liouville Equation--Classical Master Equation}

To incorporate explicit classical nuclear motion, we apply a partial Wigner transform to Equation~\eqref{eq:F-QME} with respect to the nuclear DOFs. Retaining terms to leading order in the Moyal expansion yields the F-QCLE-CME:
\begin{equation}
\begin{aligned}
\frac{\partial}{\partial t}\hat{\rho}_{sW}^F(\mathbf{R},\mathbf{P},t)
={}&
\frac{1}{2}\bigl\{\hat{H}_{sW}^F(\mathbf{R},\mathbf{P}),\hat{\rho}_{sW}^F(\mathbf{R},\mathbf{P},t)\bigr\}
- \frac{1}{2}\bigl\{\hat{\rho}_{sW}^F(\mathbf{R},\mathbf{P},t),\hat{H}_{sW}^F(\mathbf{R},\mathbf{P})\bigr\}
\\
&{}- \frac{i}{\hbar}\bigl[\hat{H}_{sW}^F(\mathbf{R},\mathbf{P}),\hat{\rho}_{sW}^F(\mathbf{R},\mathbf{P},t)\bigr]
- \hat{\hat{\mathcal{L}}}_{sb,W}^F(\mathbf{R})\,\hat{\rho}_{sW}^F(\mathbf{R},\mathbf{P},t).
\end{aligned}
\label{eq:F-QCLE-CME}
\end{equation}
Here, $\{\cdot,\cdot\}$ denotes the Poisson bracket defined in Equation~\eqref{eq:poisson_bracket}. Compared with the closed-system F-QCLE in Equation~\eqref{eq:F-QCLE}, Equation~\eqref{eq:F-QCLE-CME} contains a partially Wigner-transformed Redfield dissipator $\hat{\hat{\mathcal{L}}}_{sb,W}^F(\mathbf{R})$, which depends parametrically on the classical nuclear coordinates:
\begin{equation}
\begin{aligned}
\hat{\hat{\mathcal{L}}}_{sb,W}^F(\mathbf{R})\,
\hat{\rho}_{sW}^F(\mathbf{R},\mathbf{P},t)
={}&
\frac{1}{\hbar^2}
\int_{0}^{\infty}\! \mathrm{d}\tau\;
e^{-\tfrac{i}{\hbar}\hat{H}_{sW}^F(\mathbf{R},\mathbf{P}) t}
\mathrm{Tr}_b
\Bigl[
    \hat{H}_{sb,I,W}^F(\mathbf{R},t),
    \Bigl[
        \hat{H}_{sb,I,W}^F(\mathbf{R},t-\tau),
        \\
        &e^{\tfrac{i}{\hbar}\hat{H}_{sW}^F(\mathbf{R},\mathbf{P}) t}
        \hat{\rho}_{sW}^F(\mathbf{R},\mathbf{P},t)
        e^{-\tfrac{i}{\hbar}\hat{H}_{sW}^F(\mathbf{R},\mathbf{P}) t}
        \otimes
        \hat{\rho}_b^{\mathrm{eq}}
    \Bigr]
\Bigr]
e^{\tfrac{i}{\hbar}\hat{H}_{sW}^F(\mathbf{R},\mathbf{P}) t}.
\end{aligned}
\label{eq:LsbW-F}
\end{equation}
Here, $\hat{H}_{sb,I,W}^F(\mathbf{R},t)$ denotes the partially Wigner-transformed system--bath coupling in the interaction picture, with the nuclear coordinates entering as classical parameters.

\paragraph{Floquet Surface Hopping}

Equation~\eqref{eq:F-QCLE-CME} can be solved using a trajectory-based F-SH algorithm. As in the closed-system case, the classical nuclei evolve on a single active Floquet quasi-energy surface labeled by $N$ according to Newtonian equations of motion. Stochastic hops are introduced to reproduce the population dynamics of the Floquet electronic states. In the open-system case, however, the population change of state $M$ contains not only the standard derivative-coupling contribution but also an additional dissipative contribution from the partially Wigner-transformed bath superoperator.

Accordingly, the total hopping rate from state $N$ to state $M$ is written as the sum of a derivative-coupling-induced contribution and a bath-induced contribution:
\begin{equation}
\begin{aligned}
k_{N\rightarrow M}
&=
k_{N\rightarrow M}^{D}
+
k_{N\rightarrow M}^{\mathcal{L}}
\\
&=
\max
\left\{
0,
-2\,\mathrm{Re}
\left[
\dot{\mathbf{R}}\cdot
\mathbf{d}_{MN}^{F}(\mathbf{R})
\frac{
\sigma_{NM}^{F(\mathrm{ad})}
}{
\sigma_{NN}^{F(\mathrm{ad})}
}
\right]
\right\}
-
\bigl[
\mathcal{L}_{sb,W}^{F(\mathrm{ad})}(\mathbf{R})
\bigr]_{MM,NN}.
\end{aligned}
\label{eq:open_fsh_rate}
\end{equation}
The first term has the same origin as in closed-system F-SH and accounts for nonadiabatic transitions induced by nuclear motion. The second term arises from molecule--metal coupling and describes bath-induced population transfer between Floquet electronic states. Here we employ the secular approximation to ignore the off-diagonal part, which is reliable in long-time dynamics. The post-hop momentum rescaling follows the same procedure as in closed-system F-SH.

\subsubsection{Floquet Mixed Quantum--Classical Dynamics in the Strong-Coupling Regime}

In the strong-coupling regime ($\Gamma > k_{\mathrm{B}}T$), molecule--metal hybridization substantially broadens the molecular resonances and strongly mixes them with the metallic electronic continuum. As a result, the nuclear dynamics is more naturally described in terms of continuous electronic dissipation rather than discrete nonadiabatic hopping events. When the electronic subsystem relaxes on timescales much shorter than those of nuclear motion, its influence can be represented by effective friction and stochastic fluctuations acting on the nuclei. Under this separation of timescales, the electronic response can be expanded in the nuclear velocities, leading to an F-FPE and the corresponding F-EF algorithm.

\paragraph{Floquet Fokker--Planck Equation}

In the Markovian limit, the F-FPE describes the time evolution of the nuclear phase-space density, defined as $\mathcal{A}(\mathbf{R},\mathbf{P},t)=\mathrm{Tr}_{e,F}[\hat{\rho}_{W}^F(\mathbf{R},\mathbf{P},t)]$, where $\mathrm{Tr}_{e,F}$ denotes a trace over electronic states and Floquet indices. It takes the form
\begin{equation}
\begin{aligned}
\frac{\partial}{\partial t}\mathcal{A}(\mathbf{R},\mathbf{P},t)
={}&
-\sum_{\alpha}
    \frac{P_{\alpha}}{m_{\alpha}}
    \frac{\partial }{\partial R_{\alpha}}
    \mathcal{A}(\mathbf{R},\mathbf{P},t)
-\sum_{\alpha}
    F_{\alpha}^{F}(\mathbf{R})
    \frac{\partial }{\partial P_{\alpha}}
    \mathcal{A}(\mathbf{R},\mathbf{P},t)
\\
&{}+
\sum_{\alpha,\beta}
    \gamma_{\alpha\beta}^{F}(\mathbf{R})
    \frac{\partial}{\partial P_{\alpha}}
    \left[
        \frac{P_{\beta}}{m_{\beta}}\,
        \mathcal{A}(\mathbf{R},\mathbf{P},t)
    \right]
+
\sum_{\alpha,\beta}
    \bar{D}^{F,S}_{\alpha\beta}(\mathbf{R})
    \frac{\partial^{2}}
         {\partial P_{\alpha}\,\partial P_{\beta}}
    \mathcal{A}(\mathbf{R},\mathbf{P},t).
\end{aligned}
\label{eq:F-FPE}
\end{equation}
Here, $\alpha$ and $\beta$ label nuclear DOFs, and $m_{\alpha}$, $R_{\alpha}$, and $P_{\alpha}$ denote the corresponding nuclear mass, coordinate, and momentum, respectively. $F_{\alpha}^{F}(\mathbf{R})$ is the Floquet mean force, $\gamma_{\alpha\beta}^{F}(\mathbf{R})$ is the Floquet electronic friction tensor, and $\bar{D}^{F,S}_{\alpha\beta}(\mathbf{R})$ is the symmetric part of the Floquet momentum-space diffusion tensor associated with random force fluctuations.

\paragraph{Floquet Electronic Friction}

Equation~\eqref{eq:F-FPE} can be solved using a trajectory-based F-EF algorithm, where the nuclear dynamics is governed by the following stochastic Langevin equation:
\begin{equation}
m_{\alpha}\ddot{R}_{\alpha}
=
F_{\alpha}^{F}(\mathbf{R})
-
\sum_{\beta}
\gamma_{\alpha\beta}^{F}(\mathbf{R})\dot{R}_{\beta}
+
\delta F_{\alpha}^{F}(t).
\label{eq:FEF-Langevin}
\end{equation}
The three terms on the right-hand side are the Floquet mean force, the Floquet electronic friction force, and the random force, respectively. Within the Markovian approximation underlying Equation~\eqref{eq:F-FPE}, $\delta F_{\alpha}^{F}(t)$ is taken to be Gaussian with zero mean and correlation
\begin{equation}
\left\langle
\delta F_{\alpha}^{F}(t)\delta F_{\beta}^{F}(t')
\right\rangle
=
2\bar{D}^{F,S}_{\alpha\beta}(\mathbf{R})\delta(t-t'),
\label{eq:FEF_noise_correlation}
\end{equation}
Here, the diffusion tensor is the same $\bar{D}^{F,S}_{\alpha\beta}(\mathbf{R})$ that appears in Equation~\eqref{eq:F-FPE}, ensuring that the Langevin dynamics reproduces the momentum-space diffusion term of the F-FPE.

\section{Electron Transfer at Molecule--Metal Interfaces}

\subsection{Electron Transfer Rate Calculations}

Electron transfer at molecule--metal and molecule--semiconductor interfaces is a fundamental process in electrochemistry and substrate-mediated surface photochemistry.\cite{Lindstrom2006photoinduced} Here, we consider a periodically driven Anderson--Holstein (AH) model to illustrate how external driving can modulate the electron transfer rate (ETR) near a metal surface.\cite{Wang2023electron} We compute ETRs using F-SH and compare them with a simplified Floquet extension of Marcus theory, referred to here as F-Marcus. In this model, the driving field is characterized by an amplitude $A$ and frequency $\Omega$, while the molecule--metal coupling is characterized by the hybridization-induced level broadening $\Gamma$.

\begin{figure}[t]
\centering
\includegraphics[width=\linewidth]{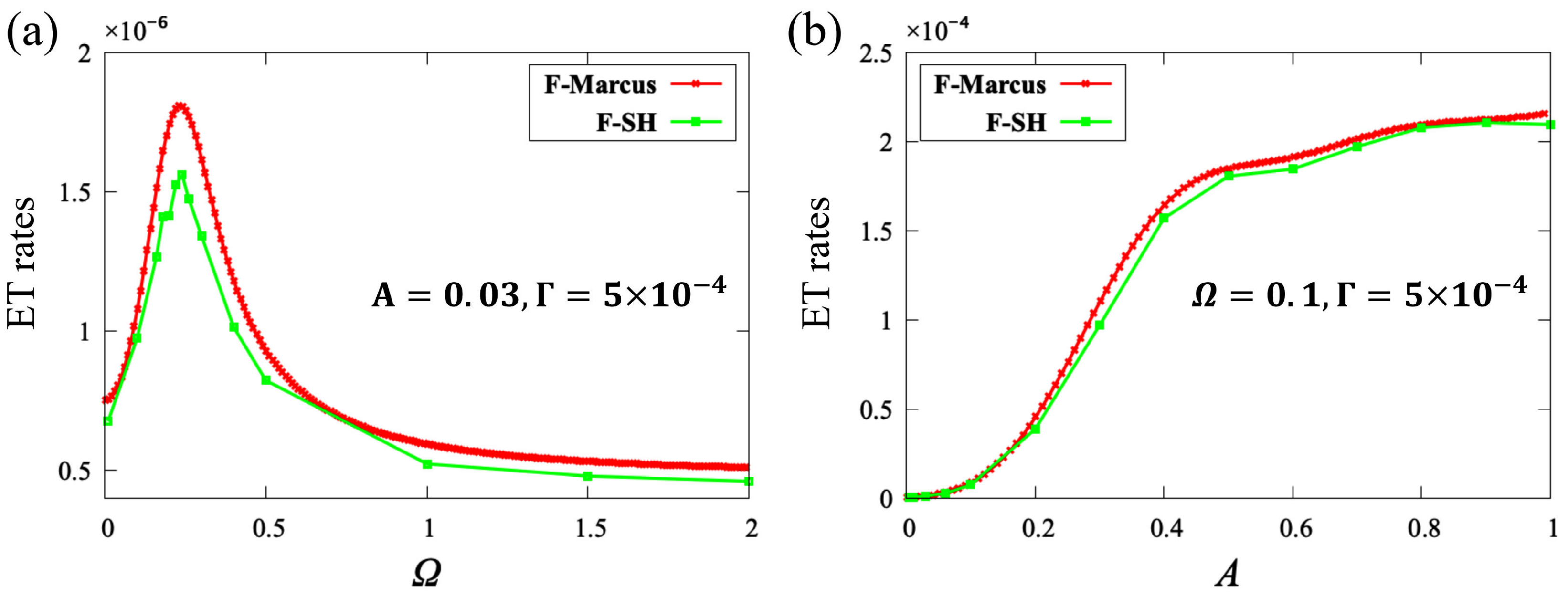}
\caption{In the weak-hybridization regime (Fermi's golden-rule limit), electron transfer rates (ETRs) obtained from Floquet surface hopping (F-SH) are compared with those from Floquet Marcus (F-Marcus) theory as functions of (a) the driving frequency $\Omega$ at fixed driving amplitude $A$ and (b) $A$ at fixed $\Omega$. The F-SH results show good agreement with the F-Marcus predictions. Adapted with permission from Reference~\citenum{Wang2023electron}. Copyright 2023 The Authors. Published by American Chemical Society.}
\label{fig:electron_transfer_rate}
\end{figure}

Figure~\ref{fig:electron_transfer_rate} summarizes how periodic driving modulates the ETR near a metal surface. In the weak-hybridization limit, represented here by $\Gamma = 0.0005$, F-Marcus agrees quantitatively with F-SH across a broad range of driving frequencies and amplitudes. Figure~\ref{fig:electron_transfer_rate}a shows a pronounced turnover of the ETR as a function of $\Omega$, with a maximum at an intermediate driving frequency and suppressed transfer in the high-frequency limit. Figure~\ref{fig:electron_transfer_rate}b shows that the ETR increases with $A$ and then saturates, without exhibiting a Marcus inverted region. This behavior reflects the role of the metallic continuum: increasing $A$ does not simply detune transfer from a single acceptor level but instead redistributes electronic weight among multiple energetically accessible Floquet sidebands.

\subsection{Nanocavity Control of Electronic Population Dynamics}

Plasmonic nanocavities can reshape local electromagnetic fields near metal surfaces, providing a route to modulate ultrafast nonadiabatic relaxation and interfacial charge transfer.\cite{Maccaferri2021recent} As a representative example, we consider a driven vibronic model of a multistate pyrazine molecule near a metal surface and compare population dynamics obtained from F-QME and F-SH.\cite{Wang2025manipulating} The model includes neutral diabatic states $|S_0\rangle$, $|S_1\rangle$, and $|S_2\rangle$, as well as an anionic state $|S_0^-\rangle$ coupled to the metal through a charge-transfer channel.

\begin{figure}[t]
\centering
\includegraphics[width=\linewidth]{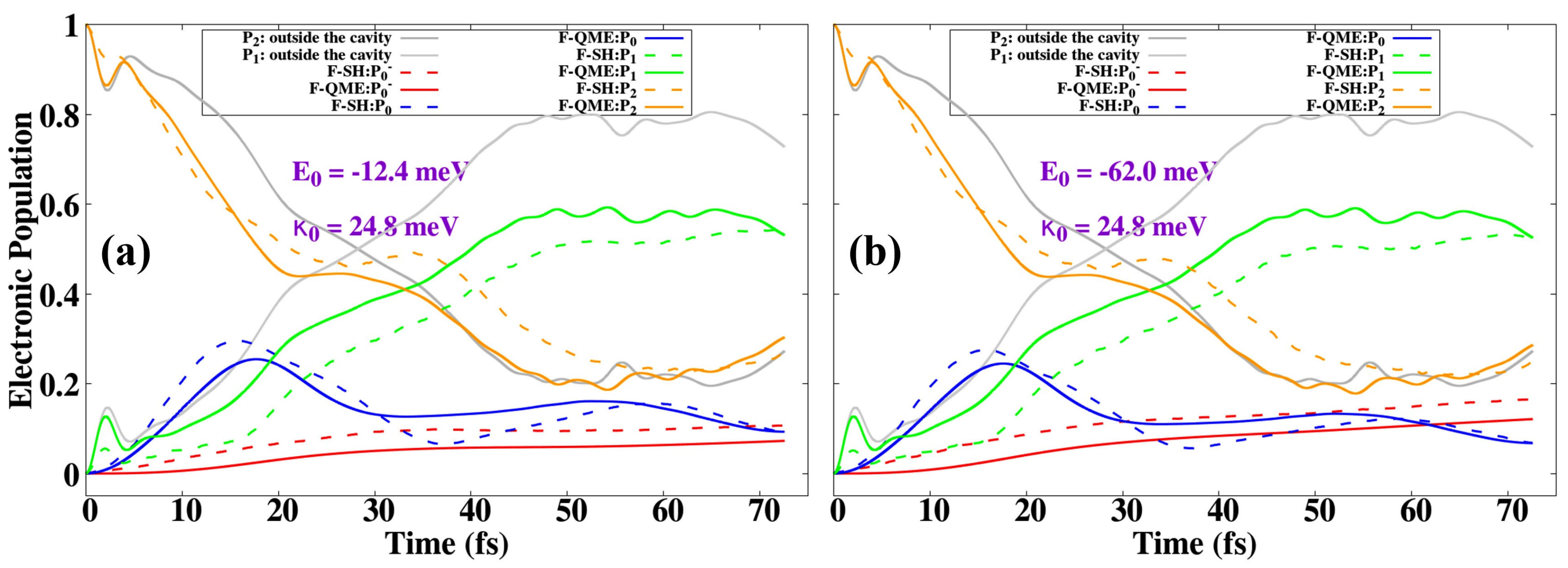}
\caption{Electronic populations in the driven pyrazine--metal model for two parameterizations of the $S_0^-$ potential energy surface (PES), characterized by different reference energy offsets $E_0$ at fixed displacement parameter $\kappa_0$. Here, $P_i$ ($i=0,1,2$) denotes the population of the neutral diabatic state $|S_i\rangle$, and $P_0^-$ denotes the population of the anionic state $|S_0^-\rangle$. Solid and dashed lines show Floquet quantum master equation (F-QME) and Floquet surface hopping (F-SH) results, respectively, while gray curves show the corresponding outside-cavity reference dynamics. Adapted with permission from Reference~\citenum{Wang2025manipulating}. Copyright 2025 American Chemical Society.}
\label{fig:pyrazine}
\end{figure}

Figure~\ref{fig:pyrazine} compares the electronic population dynamics of the driven pyrazine--metal model for two parameterizations of the $S_0^-$ PES. In both panels, the initially populated neutral excited state $|S_2\rangle$ relaxes on a femtosecond timescale, while electronic population is redistributed among the lower neutral states and gradually accumulates in the anionic state $|S_0^-\rangle$. The growth of $P_0^-$ therefore provides a direct measure of electron uptake from the metal surface. Compared with the outside-cavity reference dynamics, the cavity-modified dynamics alters the relaxation pathways and the rate of anion-state formation, demonstrating that plasmonic nanocavities can modulate both excited-state relaxation and interfacial charge transfer. The stronger stabilization of the anionic surface in Figure~\ref{fig:pyrazine}b promotes charge transfer from the metal to the molecule, as reflected by the enhanced buildup of $P_0^-$. Across both parameterizations, F-QME and F-SH yield closely consistent electronic population dynamics, supporting the reliability of the trajectory-based description in this regime.

\section{Quantum Transport in Molecular Junctions}

\subsection{Nuclear Distribution Dynamics with Photoinduced Lorentz-like Forces}

To describe light-driven nuclear dynamics in molecular junctions, we consider a spinless two-terminal model in which a molecular region is coupled to two macroscopic electrodes and driven by linearly polarized light.\cite{Chen2024floquet_lorentz} Within F-EF, periodic driving can generate a Lorentz-like force on nuclear motion through the antisymmetric, nondissipative component of the Floquet electronic friction tensor. For two-dimensional nuclear motion, this component can be characterized by
\begin{equation}
\gamma_A^F(\mathbf{R})
=
\frac{\gamma_{xy}^F(\mathbf{R})-\gamma_{yx}^F(\mathbf{R})}{2}.
\label{eq:antisymmetric_friction}
\end{equation}
Here, $\mathbf{R}=(x,y)$ denotes the nuclear coordinates. A nonzero $\gamma_A^F(\mathbf{R})$ produces a transverse force on the nuclei, analogous to a Lorentz force. Periodic driving can induce such a force even at zero bias, while nonequilibrium transport can further modulate it through the lead chemical potentials. Below, we use the symmetric bias convention $\mu_{\mathrm{R}}=-\mu_{\mathrm{L}}$.

\begin{figure}[t]
\centering
\includegraphics[width=\linewidth]{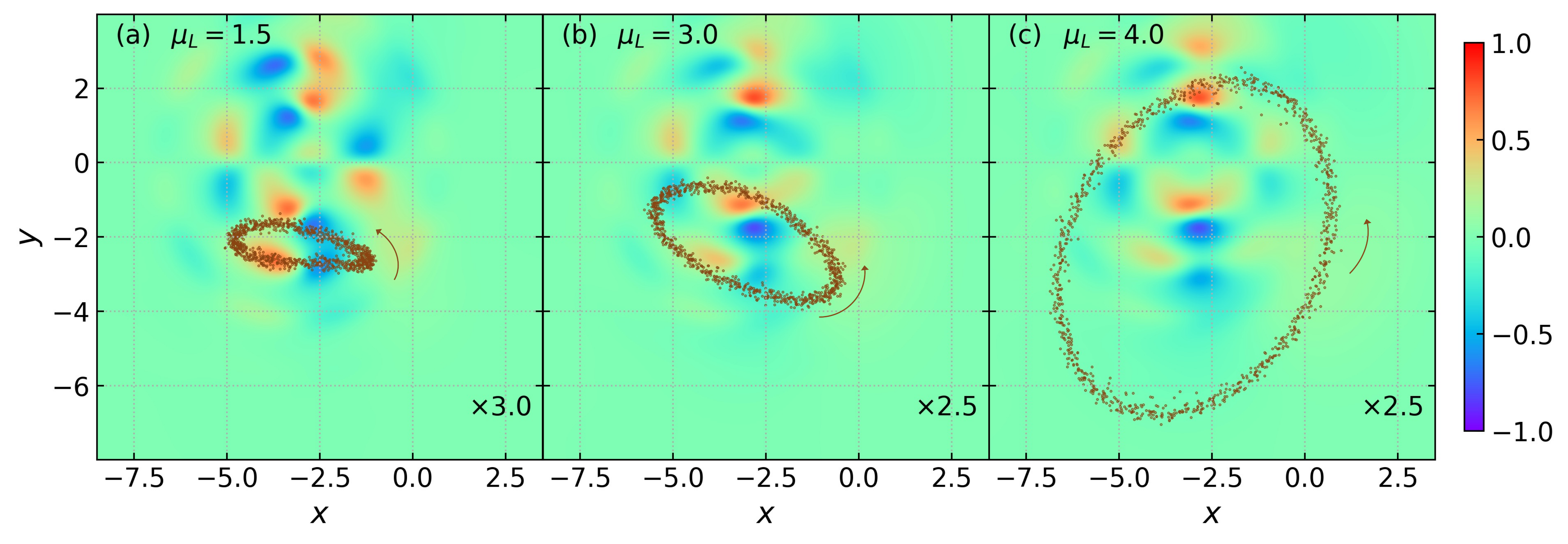}
\caption{Antisymmetric component of the Floquet electronic friction tensor, $\gamma_A^F(x,y)$, shown as color maps for different bias voltages $\mu_{\mathrm{L}}$ at fixed driving amplitude and frequency. Overlaid dots show the corresponding long-time nuclear trajectories obtained from noise-free Floquet electronic friction (F-EF) dynamics. Adapted with permission from Reference~\citenum{Chen2024floquet_lorentz}. Copyright 2024 American Chemical Society.}
\label{fig:distribution}
\end{figure}

Figure~\ref{fig:distribution} shows $\gamma_A^F(x,y)$ and the associated long-time nuclear trajectories under the same periodic driving but different bias voltages. The color maps reveal that the photoinduced antisymmetric friction is highly structured in nuclear coordinate space, with alternating positive and negative regions. Because this structure is spatially displaced from the center of the nuclear motion, the nuclei experience a net transverse force along their trajectories. As a result, the long-time dynamics forms circulating, limit-cycle-like nuclear distributions rather than simply relaxing to a static point. Increasing $\mu_{\mathrm{L}}$ modifies both the magnitude and spatial pattern of $\gamma_A^F(x,y)$, thereby changing the size and shape of the circulating nuclear trajectories. This behavior demonstrates that periodic driving and nonequilibrium transport jointly control the effective Lorentz-like force acting on nuclear motion. In this sense, light does not merely modify electronic transport directly; it can also reshape the nuclear distribution, which in turn feeds back into transport through the coordinate dependence of the molecular junction.

\subsection{Spin Polarization Dynamics Enhanced by Circularly Polarized Light}

Although the chiral-induced spin selectivity (CISS) effect has been known for decades, the experimentally observed spin polarizations are often insufficient for practical applications.\cite{Bloom2024chiral} Here we consider a chiral molecular junction driven by circularly polarized light (CPL) and examine how CPL modifies spin-dependent quantum transport.\cite{Liu2025enhancement} Using F-EF, we compute the spin-resolved currents $I^{\uparrow}$ and $I^{\downarrow}$ and quantify the spin polarization as
\begin{equation}
\xi(t)
=
\frac{I^{\uparrow}(t)-I^{\downarrow}(t)}{I^{\uparrow}(t)+I^{\downarrow}(t)}.
\end{equation}

\begin{figure}[t]
\centering
\includegraphics[width=\linewidth]{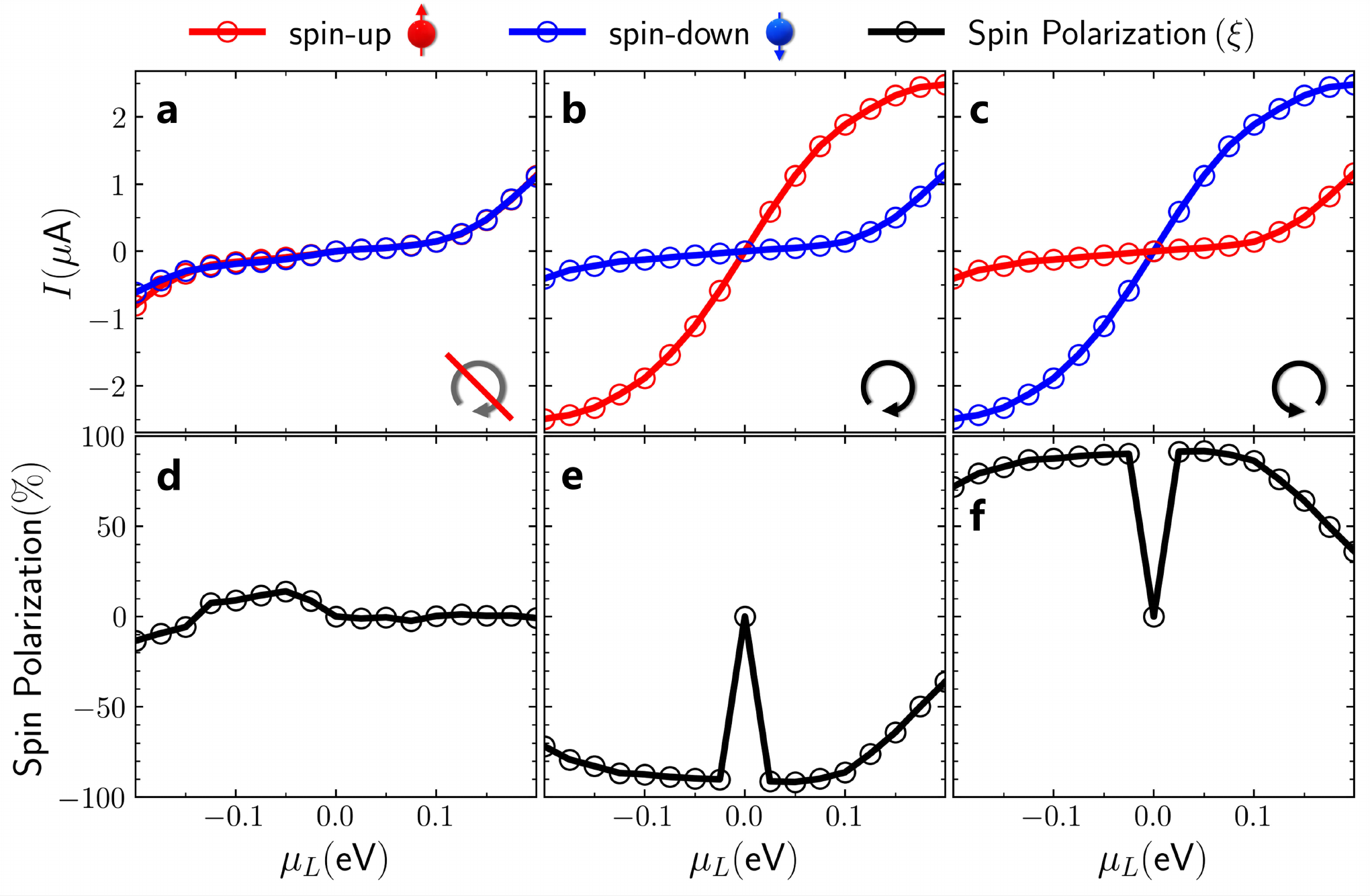}
\caption{Spin-resolved currents $I^{\uparrow}$ and $I^{\downarrow}$ (top row) and spin polarization (bottom row) as functions of $\mu_{\mathrm{L}}$ for a chiral molecular junction under the symmetric bias convention $\mu_{\mathrm{R}}=-\mu_{\mathrm{L}}$. Panels (a,d) show results without CPL, while panels (b,e) and (c,f) correspond to right- and left-handed CPL, respectively. CPL enhances the magnitude of spin polarization, and reversing the handedness switches the dominant spin channel and changes the sign of the spin polarization. Adapted with permission from Reference~\citenum{Liu2025enhancement}. Copyright 2025 American Chemical Society.}
\label{fig:SP}
\end{figure}

Figure~\ref{fig:SP} summarizes the bias dependence of the spin-resolved currents and spin polarization. The bias is swept by varying $\mu_{\mathrm{L}}$ under the symmetric convention $\mu_{\mathrm{R}}=-\mu_{\mathrm{L}}$. Without CPL, as shown in Figure~\ref{fig:SP}a,d, the two spin currents remain comparable, and spin polarization stays close to zero near zero bias. Under right-handed CPL, Figure~\ref{fig:SP}b,e, an effective light-induced Berry force promotes chiral nuclear motion and enhances the spin-up current over a finite bias window, leading to a large positive spin polarization. Reversing the handedness to left-handed CPL, Figure~\ref{fig:SP}c,f, switches the dominant spin channel, enhances the spin-down current, and yields a negative spin polarization. These results suggest that CPL provides a direct optical handle for spin polarization dynamics in chiral molecular junctions.

More recently, we have also successfully applied MF and SH methods to spin polarization dynamics in both isolated chiral molecules and open chiral molecular junctions.\cite{Han2025mixed,Wang2025mixed} Combining these methods with Floquet theory will be an important future direction for exploring more general optical-control mechanisms of the CISS effect.

\section{Carrier Dynamics in Crystalline Solids: From Real Space to Reciprocal Space}
\label{sec:k-space}

Reciprocal-space MQC formulations are particularly useful for crystalline materials because they provide direct access to band- and momentum-resolved carrier dynamics and enable Brillouin-zone truncation strategies that are difficult to implement in real space.\cite{Krotz2021a,Krotz2022a} Building on these reciprocal-space MQC formulations, we recently formulated F-MF dynamics and F-SH in real and reciprocal space.\cite{Chen2024floquet_solid}

Here, we briefly summarize how the real-space F-MF and F-SH formulations are extended to reciprocal space. For a periodic lattice with $N$ sites, a real-space variable $f_n$ can be transformed into reciprocal space as
\begin{equation}
\tilde{f}_{k}
=
\frac{1}{\sqrt{N}}
\sum_{n}
e^{ikn} f_n.
\label{eq:fourier_real_to_k}
\end{equation}
When this transformation is applied to the classical nuclear coordinates and momenta, the resulting reciprocal-space variables are denoted as $(\mathbf{R}_{\mathbf{k}}, \mathbf{P}_{\mathbf{k}})$. In this representation, the reciprocal-space Floquet electronic Hamiltonian may depend on both $\mathbf{R}_{\mathbf{k}}$ and $\mathbf{P}_{\mathbf{k}}$.

Accordingly, the real-space F-MF equations in Equation~\eqref{eq:fmf_real} become
\begin{equation}
\begin{aligned}
\dot{\mathbf{R}}_{\mathbf{k}}
={}&
\frac{\partial T_{\mathrm{nuc}}(\mathbf{P}_{\mathbf{k}})}
{\partial \mathbf{P}_{\mathbf{k}}}
+
\mathrm{Tr}
\left[
\frac{\partial \hat{H}^{F}_{\mathrm{el}}(\mathbf{R}_{\mathbf{k}},\mathbf{P}_{\mathbf{k}})}
{\partial \mathbf{P}_{\mathbf{k}}}
\hat{\rho}^{F}
\right],
\\[0.5em]
\dot{\mathbf{P}}_{\mathbf{k}}
={}&
-
\mathrm{Tr}
\left[
\frac{\partial \hat{H}^{F}_{\mathrm{el}}(\mathbf{R}_{\mathbf{k}},\mathbf{P}_{\mathbf{k}})}
{\partial \mathbf{R}_{\mathbf{k}}}
\hat{\rho}^{F}
\right].
\end{aligned}
\label{eq:fmf_k}
\end{equation}
Compared with Equation~\eqref{eq:fmf_real}, the reciprocal-space equations contain an additional electronic contribution to $\dot{\mathbf{R}}_{\mathbf{k}}$ because the Floquet electronic Hamiltonian may explicitly depend on $\mathbf{P}_{\mathbf{k}}$.

Similarly, the real-space F-SH hopping rate in Equation~\eqref{eq:fsh_real} is generalized to include Floquet derivative couplings with respect to both $\mathbf{R}_{\mathbf{k}}$ and $\mathbf{P}_{\mathbf{k}}$:
\begin{equation}
k_{N\rightarrow M}
=
\max
\left\{
0,
-2\,\mathrm{Re}
\left[
\left(
\dot{\mathbf{R}}_{\mathbf{k}}\cdot \mathbf{d}_{MN}^{\mathbf{R}_{\mathbf{k}},F}
+
\dot{\mathbf{P}}_{\mathbf{k}}\cdot \mathbf{d}_{MN}^{\mathbf{P}_{\mathbf{k}},F}
\right)
\frac{\sigma_{NM}^{F(\mathrm{ad})}}{\sigma_{NN}^{F(\mathrm{ad})}}
\right]
\right\}.
\label{eq:fsh_k}
\end{equation}
Here, $\mathbf{d}_{MN}^{\mathbf{R}_{\mathbf{k}},F}$ and $\mathbf{d}_{MN}^{\mathbf{P}_{\mathbf{k}},F}$ denote the Floquet derivative couplings with respect to $\mathbf{R}_{\mathbf{k}}$ and $\mathbf{P}_{\mathbf{k}}$, respectively. After a hop, the reciprocal-space coordinates and momenta may both need to be adjusted along their corresponding Floquet derivative-coupling directions to conserve the total energy.

Using this real- and reciprocal-space framework, we considered a driven Holstein model and compared four implementations: real-space F-MF (R-F-MF), reciprocal-space F-MF (K-F-MF), real-space F-SH (R-F-SH), and reciprocal-space F-SH (K-F-SH).

\begin{figure}[t]
\centering
\includegraphics[width=\linewidth]{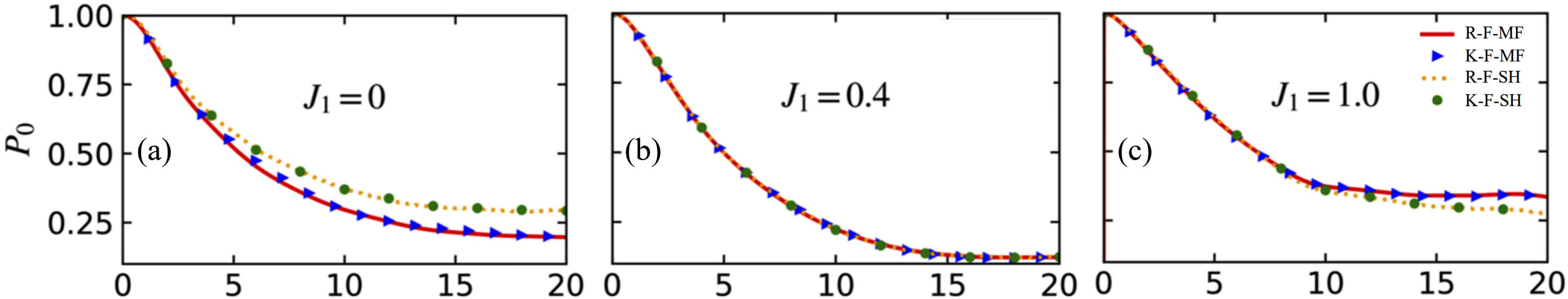}
\caption{Electronic population dynamics at $k=0$ for the driven Holstein model at different driving amplitudes $J_1$. The four curves compare real-space F-MF (R-F-MF), reciprocal-space F-MF (K-F-MF), real-space F-SH (R-F-SH), and reciprocal-space F-SH (K-F-SH). The agreement between real- and reciprocal-space results validates the consistency of the two representations, while the qualitative agreement between F-MF and F-SH suggests that the lower-cost F-MF dynamics can capture the main dynamical trends. Adapted with permission from Reference~\citenum{Chen2024floquet_solid}. Copyright 2024 AIP Publishing.}
\label{fig:r_vs_k}
\end{figure}

Figure~\ref{fig:r_vs_k} shows the electronic population $P_0$ at $k=0$ for increasing driving amplitudes $J_1$ at fixed driving frequency. The close agreement between real- and reciprocal-space results demonstrates that the two representations describe the same underlying Floquet carrier dynamics. This equivalence is important because the two formulations offer complementary advantages. Real-space simulations are conceptually simple and straightforward to implement, making them useful for local carrier motion and disordered systems. Reciprocal-space simulations, by contrast, provide direct access to band- and momentum-resolved dynamics and are therefore particularly natural for crystalline materials. Moreover, Figure~\ref{fig:r_vs_k} also highlights the different roles of F-MF and F-SH. As verified in previous studies, F-SH generally provides a more reliable description of long-time population dynamics, and F-MF remains computationally efficient and can still capture qualitative trends in large systems.\cite{Krotz2022a} Thus, the two methods are not redundant: F-SH serves as a more accurate reference, whereas F-MF offers a lower-cost approximation.

Electron--phonon and light--matter interactions play central roles in carrier relaxation, optical response, and charge transport in crystalline materials. Understanding and controlling these interactions is therefore essential for designing functional materials for optoelectronics, energy conversion, and quantum technologies.\cite{Othonos1998probing,Kilina2015light} Building on the real- and reciprocal-space Floquet MQC framework, we further studied interband transitions, intraband transitions, and charge mobility using a driven two-band model with electron--phonon coupling.\cite{Wang2024interband}

\begin{figure}[t]
\centering
\includegraphics[width=\linewidth]{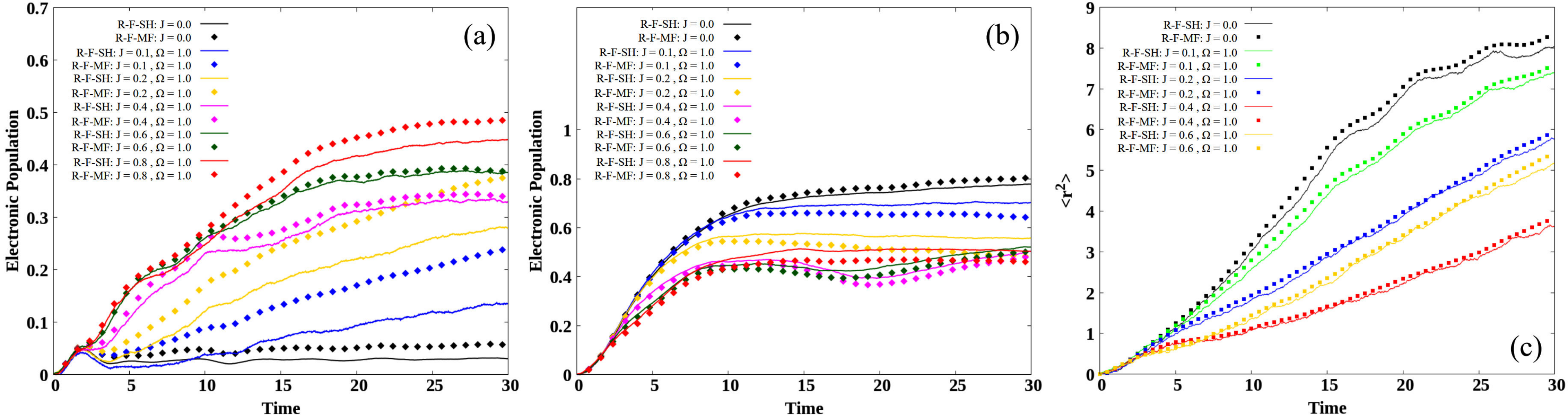}
\caption{Real-space simulations of a driven two-band model with electron--phonon coupling. Panels (a,b) compare the results of real-space F-SH (R-F-SH) and F-MF (R-F-MF) for (a) interband transition dynamics, measured by the upper-band electronic population, and (b) intraband transition dynamics, measured by the lower-band electronic population excluding the initial $k=0$ component. Panel (c) shows the mean squared displacement $\langle r^2 \rangle$ as a function of time, which serves as a dynamical measure of charge mobility. Here, $J$ is the driving amplitude and $\Omega$ is the driving frequency. Adapted with permission from Reference~\citenum{Wang2024interband}. Copyright 2024 AIP Publishing.}
\label{fig:two_band}
\end{figure}

Figure~\ref{fig:two_band} summarizes how periodic driving reshapes carrier dynamics in the two-band model. Figure~\ref{fig:two_band}a shows that increasing the driving amplitude $J$ promotes population transfer to the upper band, indicating enhanced interband transitions. In contrast, Figure~\ref{fig:two_band}b shows that the lower-band population away from the initially occupied $k=0$ component decreases with stronger driving, revealing a suppression of intraband transitions. These two trends are complementary: periodic driving opens efficient pathways for excitation across the band gap, thereby reducing the population available for redistribution within the original band. Figure~\ref{fig:two_band}c shows that the mean squared displacement $\langle r^2\rangle$ decreases as $J$ increases at fixed $\Omega$, indicating reduced spatial spreading and suppressed charge mobility. Although quantitative differences between R-F-SH and R-F-MF emerge in certain driving regimes, R-F-MF still provides a computationally efficient description that captures the main driving-induced trends. Such controllable redistribution of carrier dynamics provides a useful theoretical basis for tuning optoelectronic responses in driven semiconductors and related nanomaterials.

\section{Multicolor Floquet Engineering: From One-Frequency to Two-Frequency Driving}
\label{sec:two-frequency}

Periodic two-frequency fields, often referred to as two-color fields in experiments, provide additional control knobs beyond one-frequency driving, including independently tunable amplitudes, relative phase, polarization, and frequency detuning. These additional DOFs open new possibilities for steering nonadiabatic molecular dynamics.\cite{Il_in2014theory} Recently, we generalized the Floquet formalism from one-mode to two-mode driving by introducing two independent Fourier indices associated with the driving frequencies $\Omega_1$ and $\Omega_2$.\cite{Mosallanejad2025two} The Hamiltonian and density operator can then be expanded as
\begin{equation}
\hat{H}(t)
=
\sum_{m,n\in\mathbb{Z}}
\hat{H}^{mn}
e^{in\Omega_1 t}
e^{im\Omega_2 t},
\qquad
\hat{\rho}(t)
=
\sum_{m,n\in\mathbb{Z}}
\hat{\rho}^{mn}(t)
e^{in\Omega_1 t}
e^{im\Omega_2 t},
\label{eq:two_mode_fourier}
\end{equation}
where $\hat{H}^{mn}$ denotes the time-independent Fourier component of the two-frequency Hamiltonian, and $\hat{\rho}^{mn}(t)$ denotes the corresponding time-dependent coefficient of the density operator.

The two-mode problem can be embedded into an extended Floquet space with two independent Fourier dimensions:
\begin{equation}
\hat{H}^F
=
\sum_{m,n\in\mathbb{Z}}
\hat{L}^{\prime\prime}_{m}
\hat{L}^{\prime}_{n}
\otimes
\hat{H}^{mn}
+
\hat{N}^{\prime}
\otimes
\hat{I}\,\hbar\Omega_1
+
\hat{N}^{\prime\prime}
\otimes
\hat{I}\,\hbar\Omega_2,
\qquad
\hat{\rho}^F(t)
=
\sum_{m,n\in\mathbb{Z}}
\hat{L}^{\prime\prime}_{m}
\hat{L}^{\prime}_{n}
\otimes
\hat{\rho}^{mn}(t).
\label{eq:two_mode_HF_rhoF}
\end{equation}
Here, $\hat{L}^{\prime}_{n}$ and $\hat{N}^{\prime}$ act on the Fourier space associated with $\Omega_1$, whereas $\hat{L}^{\prime\prime}_{m}$ and $\hat{N}^{\prime\prime}$ act on the Fourier space associated with $\Omega_2$. Equation~\eqref{eq:two_mode_HF_rhoF} is the direct two-mode analogue of the one-mode Floquet representation in Equation~\eqref{eq:HF_rhoF}. With these definitions, the F-LvN equation retains the same form as Equation~\eqref{eq:flvn} and serves as the starting point for deriving two-mode extensions such as the two-mode F-QME\cite{Mosallanejad2025two} and two-mode Floquet nonequilibrium Green's function (F-NEGF).\cite{Mosallanejad2026multi}

Building on this two-mode Floquet formalism, we have recently developed a two-mode F-SH method for nonadiabatic dynamics driven by two-frequency laser fields.\cite{Han2026two} To benchmark this method, we considered a driven modified simple avoided-crossing model and used split-operator quantum dynamics as a numerically exact reference.

\begin{figure}[t]
\centering
\includegraphics[width=\linewidth]{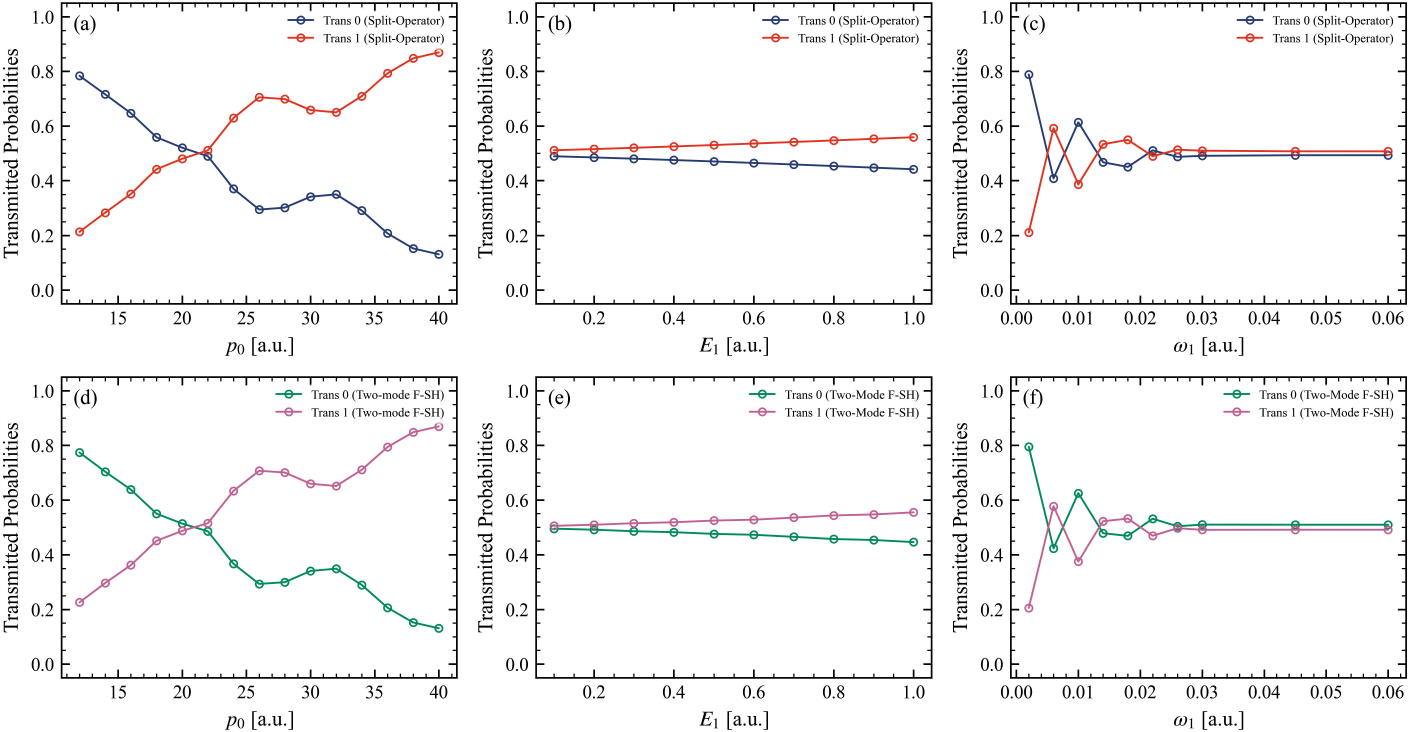}
\caption{Transmitted probabilities on the lower surface (Trans 0) and upper surface (Trans 1) for a driven modified simple avoided-crossing model. The top row shows numerically exact split-operator results, and the bottom row shows two-mode F-SH results. Panels (a,d) scan the incident nuclear momentum $p_0$, panels (b,e) scan the field amplitude $E_1$ (with $E_2 = 0.5E_1$), and panels (c,f) scan the driving frequency $\omega_1$ (with $\omega_2 = \sqrt{\omega_1}$). Adapted with permission from Reference~\citenum{Han2026two}. Copyright 2026 AIP Publishing.}
\label{fig:two_mode_fssh}
\end{figure}

Figure~\ref{fig:two_mode_fssh} compares the transmitted probabilities predicted by split-operator quantum dynamics and two-mode F-SH across different incident nuclear momenta and field parameters. The benchmark includes cases with unequal field amplitudes and noncommensurate driving frequencies, thereby testing the robustness of two-mode F-SH under general two-frequency driving conditions. The close agreement in these benchmark calculations indicates that two-mode F-SH can accurately capture nonadiabatic effects under two-frequency driving and establishes it as a practical method for simulating two-frequency control protocols. Looking forward, this two-mode framework is especially promising for studying dynamics near two-color light-induced conical intersections (LICIs). Work along this direction is currently in progress.

\section{Challenges and Opportunities}
\label{sec:challenges_and_opportunities}

Floquet nonadiabatic dynamics for light--matter interactions has advanced rapidly in recent years. Nevertheless, several conceptual and computational challenges must be overcome before these methods can be routinely applied to quantitatively reliable simulations of realistic molecular, interfacial, and materials systems.

\subsection{Scalability and Truncation}

A central computational challenge in Floquet nonadiabatic dynamics is the rapid growth of the extended Hilbert space. Achieving convergence may require many Floquet replicas, particularly under large driving amplitudes and low driving frequencies, which can substantially increase the computational cost.\cite{Wang2023nonadiabatic_FSH} Efficient truncation strategies are therefore essential for balancing computational cost against the desired accuracy. At the same time, explicitly time-dependent propagation remains an indispensable complementary option, particularly for short pulses or nonperiodic driving. Systematic benchmarks across model classes and parameter regimes are needed to delineate when Floquet-based methods provide clear advantages and when direct time-dependent methods become more appropriate.

\subsection{Nuclear Quantum Effects}

Most practical Floquet nonadiabatic simulations of long-time dynamics still rely on a classical description of nuclear motion. However, nuclear quantum effects can be significant for high-frequency vibrational modes, light atoms, and tunneling coordinates.\cite{Coffman2020modeling,Agostini2018nuclear} Path-integral-based methods, such as ring-polymer molecular dynamics (RPMD), provide practical approximate routes for incorporating these effects.\cite{Zhao2023ring,Bi2024electronic,Fu2026electrochemistry} Integrating Floquet theory with nuclear-quantum dynamics methods is therefore an important future direction, with the potential to improve simulations of light-driven processes in which zero-point motion, quantum vibrational fluctuations, and tunneling play essential roles.

\subsection{Electron--Electron Interactions}

Electron--electron interactions can qualitatively modify charge transfer, transport, and excited-state dynamics by giving rise to phenomena such as Coulomb blockade, correlation-induced level renormalization, excitonic binding, and many-body scattering.\cite{Frisenda2016transition,Bian2022chargestate} Most methods discussed in this Perspective, however, either neglect explicit electron--electron correlations or absorb them into effective single-particle descriptions. Interacting Anderson--Holstein-type models have shown that Coulomb interactions can generate additional charge-state resonances and renormalize effective electronic energetics.\cite{Dou2017born,Dou2018perspective} Recent progress in Floquet nonequilibrium Green's function (F-NEGF) approaches has begun to clarify how many-body interactions manifest in periodically driven open quantum systems.\cite{Mosallanejad2024floquet} Extending these advances to strongly correlated regimes, while simultaneously retaining coupled electron--nuclear dynamics, remains an unresolved problem.

\subsection{Non-Markovian Effects}

Realistic open molecular and interfacial systems often couple to structured environments rather than idealized memoryless baths. Energy-dependent molecule--metal hybridization, finite-bandwidth electronic continua, sharp spectral features, structured phonon environments, and slow solvent or lattice modes can all give rise to long bath correlation times. Such memory effects can limit the validity of the Born--Markov approximation commonly invoked in Redfield-type quantum master equations.\cite{Bellomo2007non} Periodic driving further enriches the spectrum through Floquet sidebands, which can open additional resonant channels and modify the lifetime of system--bath and electron--nuclear correlations. A key task is to develop Floquet nonadiabatic theories with time-nonlocal memory kernels.\cite{de_Vega2017dynamics}

\subsection{First-Principles Integration}

Predictive applications ultimately require reliable \textit{ab initio} inputs, including PESs, nonadiabatic couplings, spin--orbit couplings, electron--phonon couplings, band structures, and, for open systems, energy-dependent molecule--metal hybridization functions.\cite{Curchod2018ab,Chen2021electronic,Meng2024first} Under periodic driving, accurate and gauge-consistent light--matter coupling matrix elements are also needed, together with faithful treatments of screening, local-field effects, polarization, and geometry-dependent field enhancement.\cite{Wang2021light} Progress along these lines will be essential for moving beyond model Hamiltonians toward quantitatively predictive simulations.

\section{Summary}
\label{sec:summary}

Periodically driven nonadiabatic dynamics is central to a broad range of light--matter interaction phenomena in molecular, interfacial, and condensed-phase systems. Rather than providing an exhaustive review, this Perspective presents a concise methodological roadmap for treating Floquet nonadiabatic dynamics within a unified theoretical framework. The key idea is to combine nonadiabatic dynamics with a time-independent Floquet representation, enabling a hierarchy of reduced dynamical descriptions for both closed and open systems. We illustrate the versatility of this framework through four representative application areas: electron transfer at molecule--metal interfaces, quantum transport in molecular junctions, carrier dynamics in crystalline solids, and multicolor Floquet engineering. We hope that this Perspective will help researchers within and beyond the theoretical chemistry community navigate this rapidly developing area and identify opportunities for further methodological development. At the same time, continued advances in Floquet nonadiabatic dynamics may provide useful support and guidance for experimental efforts aimed at probing and controlling light--matter interactions.

\section*{Acknowledgments}

W.D.\ acknowledges financial support from the National Natural Science Foundation of China (Grant Nos.~22273075 and 22361142829) and the Zhejiang Provincial Natural Science Foundation (Grant No.~XHD24B0301). Y.W.\ acknowledges financial support from the National Natural Science Foundation of China (Grant No.~22403077). V.M.\ acknowledges financial support from the Summer Academy Program for International Young Scientists (Grant No.~GZWZ[2022]019). The authors acknowledge computational resources and technical support from the High-Performance Computing Center at Westlake University.

\section*{Author Biographies}

\vspace{2.0em}

\noindent
\begin{minipage}{0.20\linewidth}
\centering
\includegraphics[width=\linewidth]{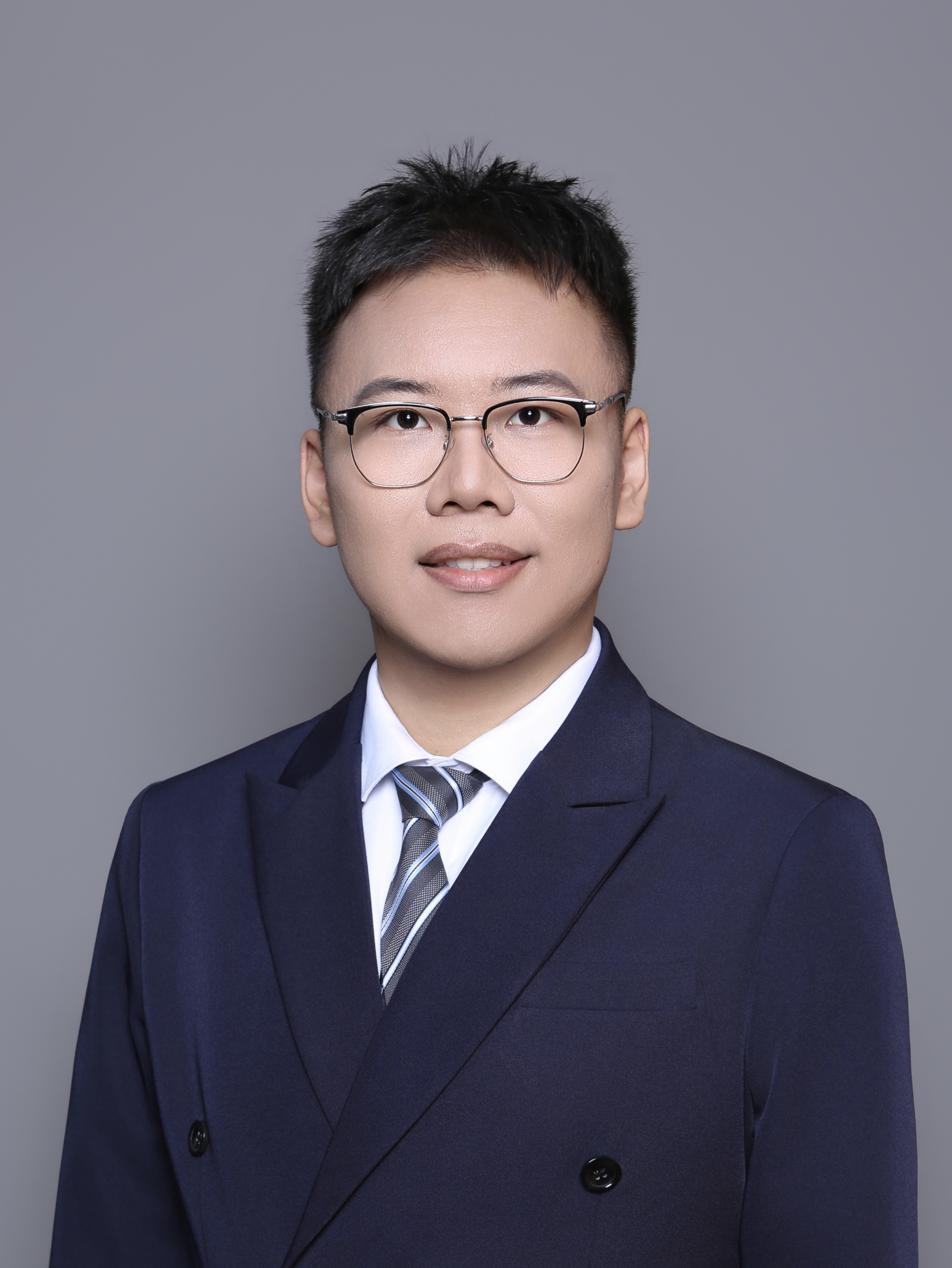}
\end{minipage}
\hfill
\begin{minipage}{0.76\linewidth}
\textbf{Jiayue Han.} Jiayue Han received his B.S.\ degree in Materials Science and Engineering from Central South University in 2022. After graduation, he worked at CNGR Advanced Material Co., Ltd., focusing on cathode materials for sodium-ion batteries. He is currently a Ph.D.\ candidate at Westlake University under the supervision of Prof.\ Wenjie Dou. His research focuses on applying nonadiabatic dynamics approaches to investigate chiral-induced spin selectivity and light--matter interactions.
\end{minipage}

\vspace{2.0em}

\noindent
\begin{minipage}{0.20\linewidth}
\centering
\includegraphics[width=\linewidth]{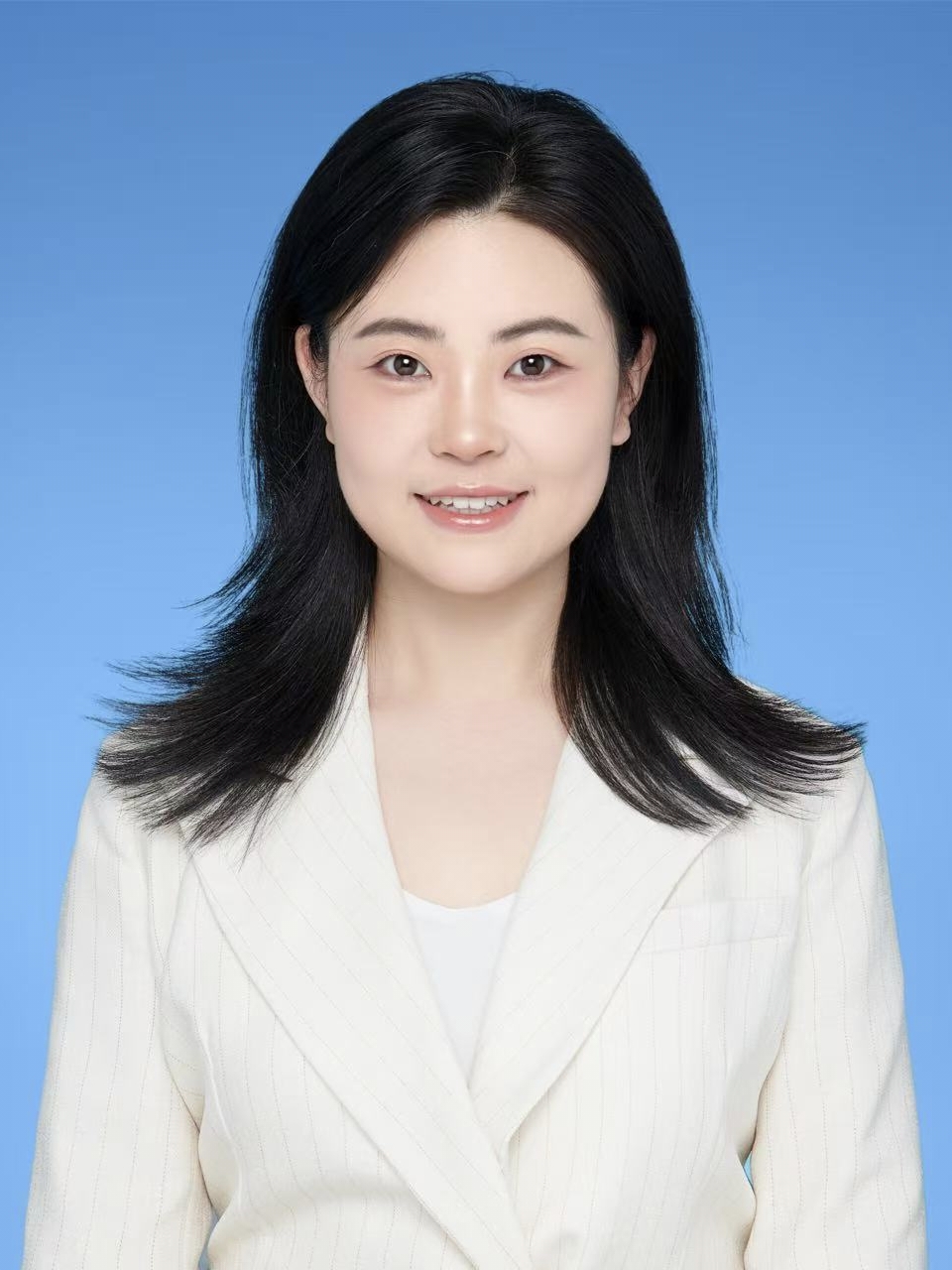}
\end{minipage}
\hfill
\begin{minipage}{0.76\linewidth}
\textbf{Yu Wang.} Yu Wang earned her B.S.\ in Chemistry from Jilin University in 2013 and her Ph.D.\ in Physical Chemistry from the same institution in 2019. She subsequently conducted postdoctoral research at Tsinghua University and Westlake University. Her research has focused on the calculation of excited-state properties of light-emitting materials, the development of nonadiabatic dynamics methods incorporating light--matter interactions, and \textit{ab initio} simulations of organic semiconductor materials. She is currently an associate professor at Hebei University of Technology.
\end{minipage}

\vspace{2.0em}

\noindent
\begin{minipage}{0.20\linewidth}
\centering
\includegraphics[width=\linewidth]{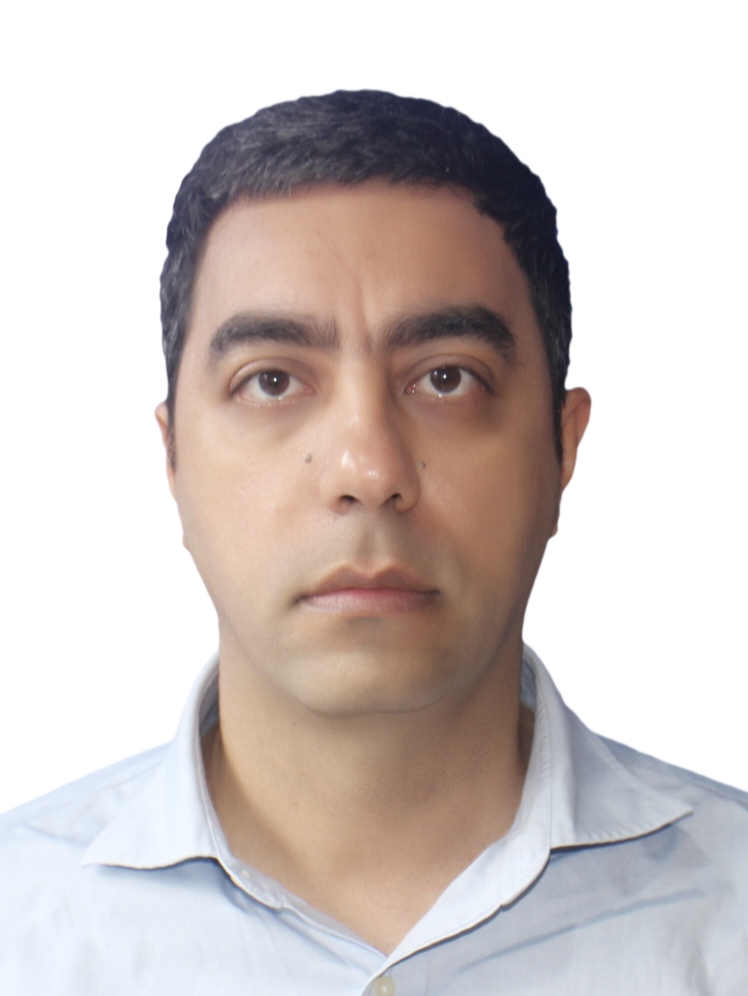}
\end{minipage}
\hfill
\begin{minipage}{0.76\linewidth}
\textbf{Vahid Mosallanejad.} Vahid Mosallanejad earned an M.Sc.\ in Solid State Physics from the University of Guilan in 2007 and an M.Sc.\ in Photonics from the Graduate University of Advanced Technology in 2011. He received a Ph.D.\ in Physics in 2017 from the University of Science and Technology of China (USTC). After completing postdoctoral training at USTC from 2017 to 2021, he joined Westlake University as a research assistant professor in chemistry. His research interests include the theory of quantum transport, light--matter interactions, and dissipation and relaxation processes in open nonequilibrium molecular systems.
\end{minipage}

\vspace{2.0em}

\noindent
\begin{minipage}{0.20\linewidth}
\centering
\includegraphics[width=\linewidth]{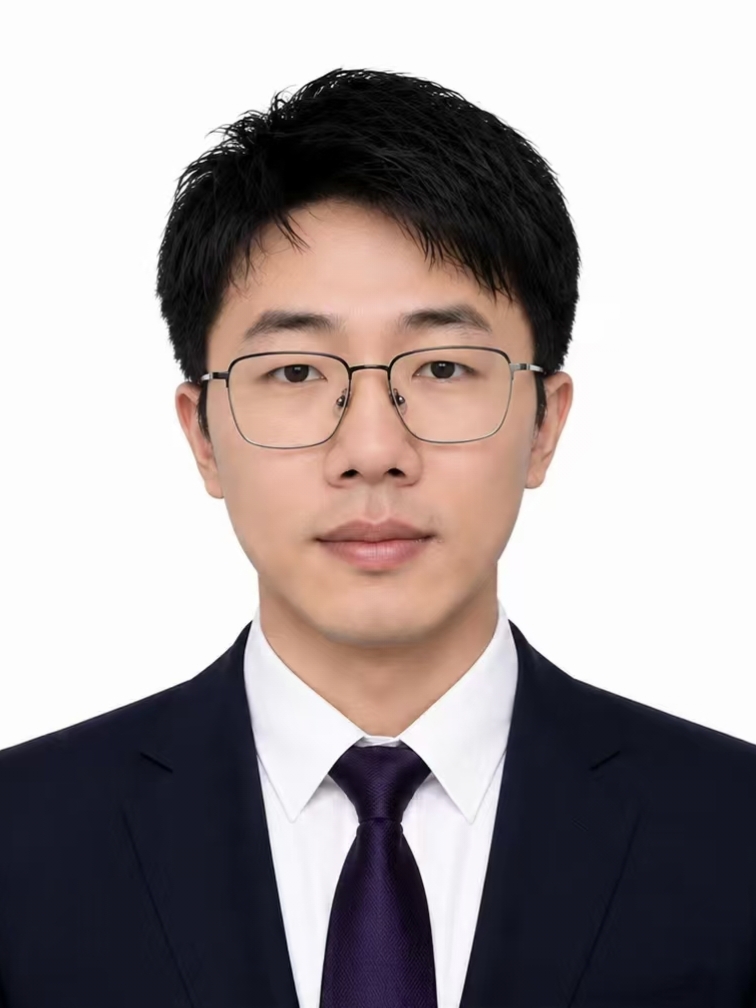}
\end{minipage}
\hfill
\begin{minipage}{0.76\linewidth}
\textbf{Wei Liu.} Wei Liu earned his B.Eng.\ (2020) and M.Eng.\ (2022) from Harbin Institute of Technology. He received his Ph.D.\ from Westlake University in 2026, supervised by Prof.\ Wenjie Dou, focusing on quantum artificial intelligence.
\end{minipage}

\vspace{2.0em}

\noindent
\begin{minipage}{0.20\linewidth}
\centering
\includegraphics[width=\linewidth]{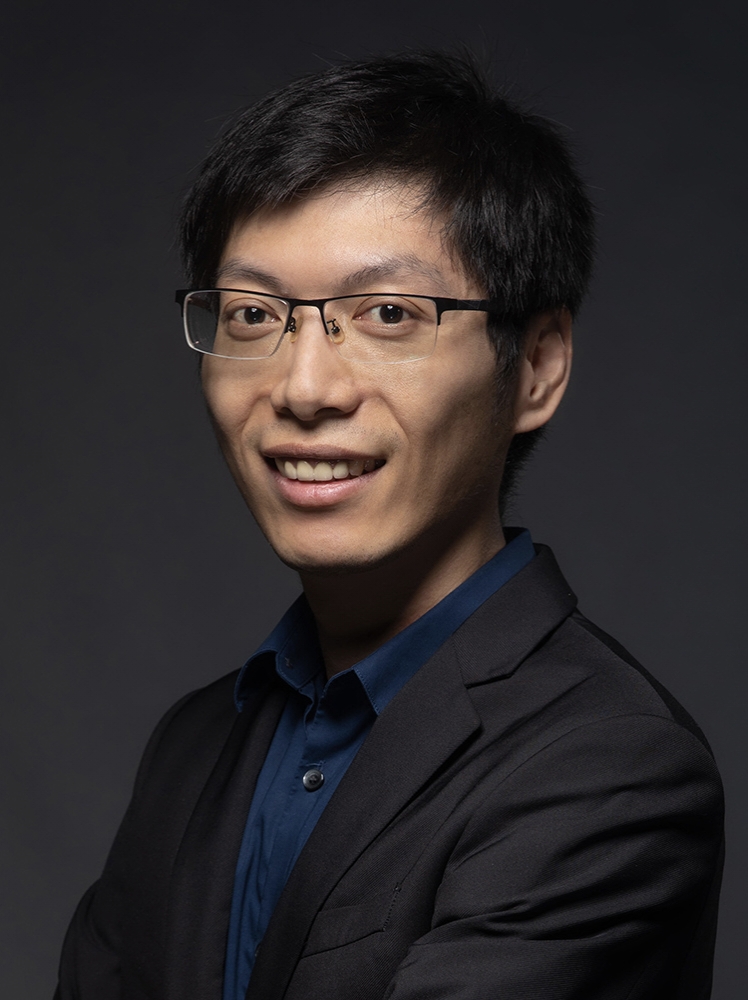}
\end{minipage}
\hfill
\begin{minipage}{0.76\linewidth}
\textbf{Wenjie Dou.} Wenjie Dou earned a B.S.\ in Physics from the University of Science and Technology of China in 2013 and a Ph.D.\ in Theoretical Chemistry from the University of Pennsylvania in 2018. His Ph.D.\ work focused on modeling nonadiabatic dynamics near surfaces. From 2018 to 2020, he was a postdoc at UC Berkeley working on stochastic implementation of electronic structure theory. He started his independent career at Westlake in January 2021. His research interests are nonadiabatic dynamics, excited-state electronic structure, and open quantum systems.
\end{minipage}

\bibliography{references}

\end{document}